\documentclass[aps,prd,nofootinbib,twocolumn]{revtex4-2}
\usepackage{amsmath,amssymb,amsfonts}
\usepackage{braket}
\usepackage{physics}
\usepackage{slashed}
\usepackage{multirow,bigdelim}
\usepackage{extarrows}

\usepackage{graphicx}
\usepackage{subfigure}

\usepackage{tikz}
\usetikzlibrary{positioning,intersections,calc,arrows.meta,math}

\usepackage{appendix}
\usepackage{color}
\usepackage{comment}

\usepackage{hyperref}

\newcommand{\imag}{\sqrt{-1}}
\newcommand{\fix}[1]{\textcolor{black}{#1}}

\begin{document}

\preprint{RIKEN-iTHEMS-25, UT-Komaba/25-10}
\title{Chiral anomaly of Kogut-Susskind fermions in the $(3+1)$-dimensional Hamiltonian formalism}

\author{Shoto Aoki}
\email{shoto.aoki@riken.jp}
\affiliation{Interdisciplinary Theoretical and Mathematical Sciences Program (iTHEMS), RIKEN, Wako 351-0198, Japan}

\author{Yoshio Kikukawa}
\email{kikukawa@hep1.c.u-tokyo.ac.jp}
\affiliation{Graduate School of Arts and Sciences, University of Tokyo, Komaba, Meguro-ku, Tokyo 153-8902, Japan}

\author{Toshinari Takemoto}
\email{takemoto@hep1.c.u-tokyo.ac.jp}
\affiliation{Graduate School of Arts and Sciences, University of Tokyo, Komaba, Meguro-ku, Tokyo 153-8902, Japan}

\begin{abstract}
We consider Kogut-Susskind fermions (also known as staggered fermions) in a $(3+1)$-dimensional Hamiltonian formalism and examine a chiral transformation and its associated chiral anomaly. The Hamiltonian of the massless Kogut-Susskind fermion has symmetry under the shift transformations in each space direction $S_k \, (k=1,2,3)$, and the product of the three shift transformations in particular (the odd shifts in general) may be regarded as a unitary discrete chiral transformation, 
modulo two-site translations. The Hermitian part of the transformation kernel $\Gamma = \imag \, S_1 S_2 S_3$ can define an axial charge as $Q_A = (1/2)\sum_x \chi^\dagger(x)  \left(\Gamma+\Gamma^\dagger \right)\chi(x)$, which is non-on site, nonquantized, and commutative with the vector charge, analogous to $\tilde{Q}_A = (1/2) \sum_n ( \chi^\dagger_n \chi_{n+1} + \chi^\dagger_{n+1} \chi_{n} )$ for the $(1+1)$-dimensional Kogut-Susskind fermion. However, our $Q_A$ cannot be expressed in terms of any quantized charges in a generalized Onsager algebra. Although $Q_A$ does not commute with the fermion Hamiltonian in general when coupled to background link gauge fields, we show that they become commutative for a class of $U(1)$ link configurations carrying nontrivial magnetic and electric fields. We then verify numerically that the vacuum expectation value of $Q_A$ satisfies the anomalous conservation law of axial charge in the continuum two-flavor theory under an adiabatic evolution of the link gauge field. 
\end{abstract}

\maketitle

\section{Introduction}

Recently, remarkable progress has been made in understanding chiral symmetry and the associated chiral anomaly of lattice fermions in the Hamiltonian formalism, particularly for Kogut-Susskind (KS) fermions~\cite{Kogut:1974ag,Banks:1975gq,Susskind:1976jm,Burden:1986by,Golterman:1984cy,Golterman:1985dz,Golterman:1986jf,Adams:2009eb,Hoelbling:2010jw,deForcrand:2010wrz,Golterman:2024xos} in $(1+1)$ dimensions~\cite{Dempsey:2022nys,Seiberg:2023cdc,Chatterjee:2024gje,Xu:2025hfs} and also in $(3+1)$ dimensions~\cite{Catterall:2025vrx, Onogi:2025xir,Gioia:2025bhl}.
In these formalisms, discrete shift symmetries under shift transformations in each spatial direction, $S_k$ $(k = 1, \dots, D)$ play an essential role in the chiral symmetry.
%The Hamiltonian of a massless KS fermion in $d=D+1$-dimensions ($D=1,3$) possesses symmetry under shift transformations in each spatial direction, $\hat{S}_k$ $(k = 1, \dots, D)$. 
Among these, the odd shifts---and in particular the diagonal shift, defined as the product of $D$ individual shifts---can be regarded as unitary discrete axial transformations, modulo two-site translations~\cite{Catterall:2025vrx}.

In $(1+1)$ dimensions, Dempsey \textit{et al.}~\cite{Dempsey:2022nys} derived a mass counterterm to improve the Hamiltonian of the massless Schwinger model, such that the shift transformation precisely induces a variation of the $\theta$ parameter by $\pi$. 
On the other hand, Shao \textit{et al.}~\cite{Chatterjee:2024gje} constructed a non-on site axial charge $Q_A$ by applying the shift transformation to a single Majorana (imaginary) component. 
This operator is conserved and quantized, does not commute with the onsite vector charge $Q_V$, and together with $Q_V$ generates an Onsager algebra~\cite{PhysRev.65.117}. 
They also introduced another non-on site axial charge, $\tilde{Q}_A$, which is conserved but not quantized, commutes with the vector charge, and still reproduces the correct axial anomaly when coupled to a $U(1)$ link field in a gauge-covariant manner.

For $(3+1)$ dimensions, Catterall \textit{et al.}~\cite{Catterall:2025vrx} extended the above analyses and introduced a non-on site, conserved, and quantized charge---analogous to $Q_A$ in $(1+1)$-dimensions---by applying the diagonal shift transformation ${S}_1 {S}_2 {S}_3$ only to a single Majorana (imaginary) component.
Furthermore, Onogi and Yamaoka~\cite{Onogi:2025xir} identified non-on site, conserved, and quantized charges $Q_{\boldsymbol{x}},(\boldsymbol{x}=\hat{1}, \hat{2},\hat{3})$, defined by acting the shift transformations ${S}_k$ $(k=1,2,3)$ on a single Majorana component.
These charges were shown to generate the $U(1)_F$ subgroup of the continuum symmetry group $SU(2)_L \times SU(2)_R \times U(1)_A$.
\footnote{For discussions of other 't~Hooft anomalies in KS fermion systems, see~\cite{Catterall:2022jky}.}
%\footnote{For the connection between Weyl fermions and Weyl semimetals, see~\cite{Armitage:2017cjs,Arouca:2022psl,Gioia:2025bhl}.}

In this paper, we further study KS fermions in the $(3+1)$-dimensional Hamiltonian formalism and examine another non-on site axial charge and its associated chiral anomaly \footnote{The chiral anomaly on the lattice space can be interpreted as the Lieb-Schultz-Mattis anomalies \cite{Cho:2017fgz, Misumi:2019jrt}}.
The Hermitian part of the unitary kernel, $\Gamma = \imag \, S_1  S_2  S_3$, of the discrete chiral transformation defines a non-on site axial charge $Q_A = (1/2) \sum_x \chi^\dagger (\Gamma + \Gamma^\dagger) \chi$, which is conserved, nonquantized, and commutes with the vector charge $Q_V$. It is analogous to $\tilde Q_A = (1/2) \sum_x (\chi^\dagger_x \chi_{x+1} + \chi^\dagger_{x+1} \chi_x)$ for the $(1+1)$-dimensional KS fermion~\cite{Chatterjee:2024gje,Catterall:2025vrx}, but, as we will see bellow, it is not related to any quantized charges of the generalized Onsager algebra in $(3+1)$ dimensions~\cite{Pace:2025rfu}, in contrast to the $(1+1)$-dimensional case where $\tilde Q_A = (1/2)(Q_1 + Q_{-1})$\footnote{
It differs from any of the two axial charges introduced by 
Catterall \textit{et al.}, but it anticommutes with the time-reversal transformation \cite{Catterall:2025vrx}.}.
We note that $\Gamma = \imag \,  S_1 S_2 S_3$ is rather related to the chiral operator $\Gamma_5$ in the Euclidean formulation~\cite{Golterman:1984cy,Golterman:1985dz,Golterman:1986jf,Adams:2009eb, Hoelbling:2010jw,deForcrand:2010wrz}.
%We note that $\Gamma = \imag \,  S_1 S_2 S_3$, because $\gamma_5 \otimes {}^t \gamma_5 = \pm 1$ in the $(3+1)$-dimensional Hamiltonian formalism, and it is rather related to the chiral operator $\Gamma_5$ in the Euclidean formulation~\cite{Golterman:1984cy,Golterman:1985dz,Golterman:1986jf,Adams:2009eb, Hoelbling:2010jw,deForcrand:2010wrz}.

Although $Q_A$ does not generally commute with the fermion Hamiltonian when coupled to a background link gauge field, we show that they become commutative for a particular class of $U(1)$ link configurations carrying nontrivial magnetic and electric fields.
We then verify numerically, with controlled accuracy, that the vacuum expectation value of $Q_A$ satisfies the anomalous axial charge conservation law of the continuum two-flavor theory~\cite{Adler:1969gk,Bell:1969ts,Dunne:1989gp,Arouca:2022psl} under adiabatic evolution of the link gauge field.
\footnote{For related studies using Wilson and overlap fermions in the Hamiltonian formalism, see \cite{Ambjorn:1983hp} and \cite{Hayata:2023zuk,Hidaka:2025ram, Horvath:1998gq,Cheluvaraja:2000an,Creutz:2001wp}.}

This paper is organized as follows. In Sec.~\ref{sec: Free KS fermion}, we review the symmetries of the KS fermion system and define the diagonal shift operator corresponding to the chiral symmetry. We also construct the generalized Onsager algebra \cite{Pace:2025rfu} associated with all shift operators.
In Sec.~\ref{sec: Anomaly Eq}, we study the vacuum expectation value of the chiral charge and demonstrate that it satisfies the chiral anomaly equation.
Finally, we summarize our results and present concluding remarks in Sec.~\ref{sec: Conclusion}.

\section{Free KS fermion Hamiltonian and Symmetries in $(3+1)$-Dimensional Spacetime}
\label{sec: Free KS fermion}

%この章では、$(d+1)$次元の時空上のKS fermionを考える。ここで時間が連続で、空間は離散であるとする。

We consider KS fermion systems in $(3+1)$-dimensional spacetime. We are interested in the case where the only time is continuous, but the space is discrete. In this section, we review the Hamiltonian of the KS fermion system and symmetries. We also define the chiral transformation in the lattice \cite{Catterall:2025vrx}. The analysis on the taste basis is discussed in Appendix \ref{sec: taste basis}.

We denote the site on the three-dimensional lattice space by
\begin{align}
    x=\qty(x^1, x^2, x^3),~ x^i=0,\cdots, N-1,
\end{align}
where $N$ is an even integer so that lattice translations do not violate periodic boundary conditions $x^i+N \sim x^i$. The Hamiltonian of the KS fermion is defined by \cite{Kogut:1974ag}
\begin{align}
    H= \sum_{x} \chi(x)^\dagger h\chi(x),~ h= \sum_{i=1}^3  \frac{ T_i -T_i^\dagger  }{2 \imag}
    + m_\text{lat}\epsilon(x)   \label{eq: Hamiltonian}
\end{align}
where $\eta_i(x)= (-1)^{x^1 +\cdots+ x^{i-1}} ,\epsilon(x)= (-1)^{x^1 +x^2+ x^{3}} $ and $T_i=\eta_i(x) \delta_{x+\hat{i},y}$ is a shift operator in the $i$ direction, 
\begin{align}
    T_i\chi(x)=\sum_{y}  (T_i)_{xy} \chi(y)=\eta_i(x) \chi(x+ \hat{i}),
\end{align}
where $\hat{i}$ represents a unit vector in the $i$ th direction. The commutation relations are given by
\begin{align}
    T_i T_j +T_j T_i= 2 \delta_{ij }T_j^2.
\end{align}

%$U_i(r)$ is a $U(1)$ link variable, which is related to a gauge field in the continuum limit. In the rest of this section, we consider the free theory.

\if0
The KS fermion field $\chi$ satisfies the periodic boundary condition:
\begin{align}
    \chi(x+N\hat{i})= \chi(x)
\end{align}
\fi

\subsection{Parity, time teversal, and charge conjugation}

The free massive KS fermion system has discrete symmetries such as parity, time reversal, and charge conjugation. We define the parity operator as \cite{Golterman:1984cy,Golterman:2024xos}
\begin{align}
    P^{-1} \chi(x) P= \epsilon( x) \chi(-x).\label{eq: parity}
\end{align}
It converges to $(\beta\otimes 1) \psi(-r)$ on the taste basis in the continuum limit (see Appendix \ref{sec: taste basis}).
\if0
on the taste basis, the parity operator acts on $\psi$ as
\begin{align}
    P^{-1}\psi(r)P= (\beta \gamma_5 \otimes 1)\Gamma^{-1} \psi(-r).
\end{align}
The right-hand side converges to $(\beta\otimes 1) \psi(-r)$ in the continuum limit.
\fi

Time reversal is defined as \footnote{There is another time reversal symmetry as
\begin{align}
    T^{\prime -1} \chi(x)T^\prime= \epsilon(x) \chi(x),~ T^{\prime -1} \imag T^\prime=-\imag,\label{eq: time reversal}
\end{align}
which acts on $\psi$ as
\begin{align}
     T^{\prime -1} \psi(r)T^\prime= T_2 S_2^{-1} \psi(r). 
\end{align}
The continuum limit is given by $-\alpha_1 \alpha_3 \otimes {}^t \sigma_2$. 
}
\begin{align}
    T^{-1}\chi(x) T&= (-1)^{x^2}T_1 T_3 \chi(x)=-(-1)^{x^1}\chi(x+\hat{1}+\hat{3}),~ \nonumber \\
    T^{-1}\imag T&= -\imag ,
\end{align}
and commutes with the Hamiltonian. In the continuum limit, $\psi(r)$ changes to $ (\alpha_1 \alpha_3 \otimes 1) \psi(r)$ on the taste basis.
\if0
on the taste basis, this operator acts on $u$ as
\begin{align}
    T^{-1}u(r) T= \frac{1}{2}\sum_{A} T^{-1}\chi(2r+A)T (\sigma^{(A)})^\ast=\frac{1}{2}\sum_{A} T_1T_3\chi(2r+A) \sigma^{(A)}=T_1T_3 u(r).
\end{align}
Then, $\psi$ changes to
\begin{align}
    T^{-1}\psi(r) T= T_1T_3 \psi(r) \to (\alpha_1 \alpha_3 \otimes 1) \psi(r).
\end{align}
This result is consistent with the time reversal in continuum space.
\fi

We also find charge conjugation,
\begin{align}
    C^{-1} \chi(x) C= (-1)^{x^2} \epsilon T_2 \chi^\ast (x)=(-1)^{x^3}\chi^\ast (x+ \hat{2}). \label{eq: charge conjugation}
\end{align}
This transformation corresponds to $ \psi(r)\to (\beta \alpha_2\otimes 1) \psi^\ast(r)$.
\if0
This transformation can be translated into the taste basis as
\begin{align}
    C^{-1}\psi(r) C= (\beta\otimes 1) T_2 \psi^\ast(r) \to (\beta \alpha_2\otimes 1) \psi^\ast(r).
\end{align}
\fi

\subsection{Shift symmetries as axial flavor and axial $U(1)$ symmetries}
The massless KS fermion system has shift symmetries. We define other shift operators by a single site along the $x^i$ direction by
\begin{align}
    S_i \chi(x)=  \xi_i(x) \chi(x+ \hat{i}), \label{eq: axial shift}
\end{align}
where $\xi_i(x)=(-1)^{x^{i+1} +\cdots +x^3}$ \cite{Golterman:1984cy, Catterall:2025vrx}. The commutation relations are given by
\begin{align}
    S_i S_j+ S_j  S_i =2 \delta_{ij} S_j^2,~S_iT_j -T_j S_i=0,~  S_i \epsilon + \epsilon S_i=0.
\end{align}
\if0
They commute with the massless Hamiltonian and can be rewritten as
\begin{align}
    S_i \psi(r)= 
    \imag (\gamma^5 \otimes {}^t \sigma_i) \psi(x)  +\imag  \frac{1}{2}(\gamma^5 \otimes {}^t \sigma_i - \beta \alpha_i\otimes 1) \nabla_i \psi(x)
\end{align}
on the taste basis. \fi
Taking the continuum limit, this operator converges to $\gamma^5 \otimes {}^t\sigma_i$ on the taste basis, which is the generator of the axial $SU(2)$ rotation with $\frac{\pi}{2}$ (see Appendix \ref{sec: taste basis}). Thus, the shift operators can be taken to be equivalent to axial $SU(2)$ rotations in the lattice space.

We also define the diagonal shift operator as \footnote{This operator is odd under Catterall's $\mathcal{T}$ transformation \cite{Catterall:2025vrx}.}
\begin{align}
    \Gamma \chi(x)=& \imag S_1 S_2 S_3 \chi(x) \nonumber\\
    =&-\imag T_1 T_2 T_3\chi(x) \nonumber \\
    =&\imag (-1)^{x^2} \chi(x+T) \label{eq: chiral operator},
\end{align}
where $T= \hat{1}+\hat{2}+\hat{3}$. $\Gamma$ commutes with the massless Hamiltonian and satisfies the lattice continuity equation \eqref{eq: conservation law j_A}. In the continuum limit, this operator corresponds to $\gamma_5 \otimes 1$ on the taste basis. Thus, we regard $\Gamma$ as the chiral operator in the lattice space. Note that $\Gamma$ is a unitary rather than a Hermitian. The (lattice) regularized chiral charge and its density are defined as
\begin{align}
    Q_{\text{reg}}=&\sum_x j_\text{reg}^0(x)= \sum_x\chi^\dagger(x) \Gamma \chi(x) \label{eq: reg chiral charge} ,~\nonumber\\
    j_\text{reg}^0(x)=&\chi^\dagger (x )\Gamma \chi(x).
\end{align}
We denote its Hermitian and anti-Hermitian parts by $Q_A$ and $\tilde{Q}$ as 
\begin{align}
    Q_A=&\sum_x j_A^0(x)= \sum_x\chi^\dagger(x) \frac{\Gamma+\Gamma^\dagger}{2} \chi(x) \label{eq: chiral charge}   ,~\\
    j_A^0(x)=&\frac{1}{2} \imag (-1)^{x^2} (\chi^\dagger(x) \chi(x+T)-  \chi^\dagger(x+T) \chi(x) ), \\
    \tilde{Q}= &\sum_x \chi^\dagger (x) \frac{\Gamma- \Gamma^\dagger }{2 \imag} \chi(x)=\sum_x  \tilde{j}^0(x) ,~\\
    \tilde{j}^0(x)=&\frac{1}{2} (-1)^{x^2} (\chi^\dagger(x) \chi(x+T)+  \chi^\dagger(x+T) \chi(x) ).% \label{eq: Onsager charge density}.
\end{align}
%This construction is similar to that of the unquantized axial charge in $(1+1)$ dimensional case \cite{Chatterjee:2024gje}. 

\fix{
As we will see below, the axial anomaly arises from the Hermitian part $Q_A$, rather than from the anti-Hermitian part $\tilde{Q}$, and $Q_A$ is essentially different from the non-quantized charge discussed in \cite{Chatterjee:2024gje}.
}

\subsection{Onsager algebra}
\label{sec: Onsager Algebra}

We consider the massless Hamiltonian and investigate algebraic relations between conserved charges with integer eigenvalues associated with shift operators \eqref{eq: axial shift}. We show that these charges generate the generalized Onsager algebra \cite{PhysRev.65.117, Pace:2025rfu}.

%We construct the Onsager algebra \cite{PhysRev.65.117} generated by these charges based on the method proposed in \cite{Catterall:2025vrx, Onogi:2025xir}.

%These charges are generators of the Onsager algebra associated with the diagonal shift operator \eqref{eq: chiral operator}. 

In the massless case, there is an on-site charge conjugation,
\begin{align}
    \chi(x) \mapsto \chi^\ast (x).
\end{align}
This transformation leads to 
\begin{align}
    \psi(r) \mapsto (\beta\otimes 1) T_2 S_2^{-1} \psi(r)
\end{align}
on the taste basis. In the method proposed in the previous works \cite{Chatterjee:2024gje, Catterall:2025vrx, Onogi:2025xir}, the staggered fermion $\chi$ is decomposed into two Majorana fermions under the on-site charge conjugation as %\footnote{The real fermion fields $a$ and $b$ are not Majorana fermions under our charge conjugation.}
\begin{align}
    \chi(x)= \frac{1}{2}( a(x)+ \imag b(x)) , \label{eq: majorana rep}
\end{align}
where 
\begin{align}
    \qty{a(x), a(x^\prime)}= \qty{b(x), b(x^\prime)}=2\delta_{xx^\prime},~\qty{a(x), b(x^\prime)}=0.
\end{align}
Then, the massless Hamiltonian is given by
\begin{align}
    H= \frac{\imag}{4} \sum_{x} \eta_i(r) \qty( a(x)a(x+\hat{i}) + b(x)b(x+\hat{i}) ).
\end{align}

As we saw before, this Hamiltonian has shift symmetries under $S_i$. These operators generate a symmetry group
\begin{align}
    G=\langle -I, S_1,S_2,S_3 \mid S_iS_j= -S_j S_i ~(i\neq j)\rangle,
\end{align}
where $I$ is an identity element. Note that $G$ is a central extension of $\mathbb{Z}^3$ by $\mathbb{Z}_2$. The translational operator $\hat{S}_i^{(b)}$ that acts on only $b(x)$ in the $i$ th direction is given by
\begin{align}
   \hat{S}_i^{(b)} a(x) (\hat{S}_i^{(b)})^{-1} &=a(x),~\nonumber \\
   \hat{S}_i^{(b)}b(x) (\hat{S}_i^{(b)})^{-1} &=S_ib(x)=\xi_i(x)b(x+ \hat{i}).
\end{align}

We extend this representation to any group element in $G$. If $g \in G$ is written as $g= S_{g_1} S_{g_2}  \cdots S_{g_n}  ~(g_i=1,2,3)$, we can define
\begin{align}
    g(x)&= x+ \sum_{i=1}^n \hat{g}_i,~\\
    \xi_g(x)&= \xi_{g_1}(x+  \sum_{i=2}^{n} \hat{g_i} ) \xi_{g_2}(x+  \sum_{i=3}^{n} \hat{g_i} ) \cdots   \xi_{g_n}(x)
\end{align}
and the translational operator $\hat{S}_g^{(b)}$ as
\begin{align}
    \hat{S}_g^{(b)}b(x) (\hat{S}_g^{(b)})^{-1} =&\hat{S}_{g_1}^{(b)}\cdots  \hat{S}_{g_n}^{(b)}b(x) (\hat{S}_{g_n}^{(b)})^{-1} \cdots (\hat{S}_{g_1}^{(b)})^{-1} \nonumber \\
    =&S_{g_n} \cdots S_{g_1}b(x)= b(g(x)) \xi_g(x).
\end{align}
We assume that $g \mapsto (\hat{S}_g^{(b)})$ is a group homomorphism. Then, $\xi$ should satisfy
\begin{align}
    \xi_{\pm I}(x)=\pm1,~\xi_{gh}(x)=\xi_g(h(x)) \xi_h(x)
\end{align}
for $g, h \in G$, and 
\begin{align}
    \xi_{g^{-1}} (x)=\xi_g(g^{-1}(x))
\end{align}
for the inverse element of $g$. Here, we use $\xi_g(x)=\pm1$.

We define the quantized and conserved charges as
\begin{align}
Q_I=\sum_x \qty[\chi^\dagger (x) \chi(x)-\frac{1}{2}]=\frac{\imag}{2} \sum_x a(x) b(x),\\
    Q_g=  \hat{S}_g^{(b)} Q_I (\hat{S}_g^{(b)})^{-1} =\frac{\imag}{2} \sum_x \xi_g(x) a(x)b(g(x)) 
\end{align}
and the auxiliary generator as
\begin{align}
    G_{g,h}= \frac{\imag}{2} \sum_x \mqty[ \xi_g(x) a(x)a(g(x))- \xi_h(x) b(x)b(h(x))].
\end{align}
$Q_I$ is a vector $U(1)$ charge whose eigenvalue expresses the fermion number \footnote{$Q_I$ is equivalent to $Q_V$ or $Q_0$ in \cite{Chatterjee:2024gje, Pace:2025rfu, Onogi:2025xir, Catterall:2025vrx}.}.
These operators commute with the massless Hamiltonian and satisfy the $G$-Onsager algebra relations \cite{Pace:2025rfu},
\begin{align}
    [Q_g, Q_h]=&\imag G_{g^{-1}h ,hg^{-1} },\\
    [Q_g, G_{h,k}]=&\imag( Q_{gh^{-1}} + Q_{k^{-1}g} - Q_{gh}-Q_{kg}),\\
    [G_{g_1,h_1}, G_{g_2,h_2}]=& \imag (G_{g_2g_1, h_1h_2} - G_{g_2^{-1}g_1, h_1h_2^{-1}} \nonumber  \\
    &- G_{g_1g_2, h_2h_1} + G_{g_1g_2^{-1}, h_2^{-1}h_1} ) .
\end{align}
We denote the infinite-dimensional Lie algebra $\qty{Q_g, G_{h,k}}$ by $\mathbf{Ons}_3$ \footnote{$\mathbf{Ons}_3$ contains Onsager sub-algebras, such as those generated by $Q_I$ and $Q_{S_T}$ discussed in \cite{Catterall:2025vrx}, and those generated by $Q_I$ and $Q_{S_3}$ discussed in \cite{Onogi:2025xir}.
%The Onsager algebra discussed in \cite{Onogi:2025xir} is a sub-algebra of $\mathbf{Ons}_3$, generated by $Q_I$ and $Q_{S_3}$.
}.

The regularized chiral charge $Q_\text{reg}$ belongs to $\mathbf{Ons}_3$. Setting $S_T= S_3 S_2 S_1$, we can rewrite $Q_\text{reg}$ as
\begin{align}
    Q_\text{reg}=&Q_A +\imag \tilde{Q} \nonumber \\ 
    =  &\frac{\imag}{4} \sum_x (-1)^{x^2} (a(x) a(x+T) +b(x )b(x+T)\nonumber \\
    &+ \imag( a(x) b(x+T) -a(x)b(x-T) ) ) \nonumber \\
    = &\frac{1}{2} G_{S_T,-S_T} + \imag \frac{ Q_{S_T} +Q_{S_T^{-1}} }{2} .
\end{align}
Since $S_T$ is a central element of $G$, $Q_A=\frac{1}{2}G_{S_T,-S_T}$ commutes with all elements of $\mathbf{Ons}_3$. That is, the Hermitian chiral operator $Q_A$ \footnote{\fix{Our $Q_A$ is not quantized and is consistent with the Nielsen-Ninomiya theorem \cite{Nielsen:1981hk}. However, this does not mean that there is no well-defined definition on the lattice. }
} can be interpreted as the central charge of the KS fermion. %\fix{As we will see below, the axial anomaly arises from the Hermitian part $Q_A$ rather than the anti-Hermitian part $\tilde{Q}$, and $Q_A$ is essentially different from the non-quantized charge discussed in \cite{Chatterjee:2024gje}}

On the other hand, $Q_{S_T}$ acts on two Majorana fields as
\begin{align}
    [Q_{S_T}, a(x)]&= -\imag (-1)^{x^2}b(x+T), \\
    [Q_{S_T}, b(x)]&= -\imag (-1)^{x^2}a(x-T).
\end{align}
These equations lead to
\begin{align}
    [Q_{S_T}, \psi(r)] \to -\imag (\beta \alpha_2\otimes {}^t \sigma_2) \psi^\ast(r)
\end{align}
in the continuum limit \cite{Onogi:2025xir}. This transformation looks like the charge conjugation \eqref{eq: charge conjugation} with a flavor rotation rather than the chiral transformation. The derivation is implemented in Appendix \ref{sec: taste basis}. \fix{
Our $Q_{S_T}$ is constructed from the translation of the Majorana $b$ field, in a manner similar to that in \cite{Chatterjee:2024gje}. However, the diagonal translation operator $S_T$ cannot be interpreted as $\mathcal{C}^R$, which corresponds to the charge conjugation of right-moving particles in $(1+1)$ dimensions. In fact, while the commutation relation of $\psi$ resembles that of charge conjugation, the quantized charge in $(1+1)$ dimensions is associated with the axial transformation. Therefore, although the construction is formally similar to that in \cite{Chatterjee:2024gje}, its physical interpretation is different.
}
%\fix{Our $Q_{S_T}$ is constructed by the translation for the Majorana $b$ field as well as \cite{Chatterjee:2024gje}. However, the diagonal translational operator $S_{T}$ is not interpreted as $\mathcal{C}^R$, which is the charge conjugation for right-moving particles in $(1+1)$-dimension. Actually, the commutation between $\psi$ resembles charge conjugation, while the quantized charge in $(1+1)$-dimension is related to the axial charge. Although the construction is similar to \cite{Chatterjee:2024gje}, the physical meaning is different.}

%\fix{Note that our $Q_{\text{reg}}$ is essentially different from the }

\if
A trivial conserved charge is identified as a vector $U(1)$ charge, 
\begin{align}
    Q_0= \sum \qty[\chi^\dagger (x) \chi(x)-\frac{1}{2}]=\frac{\imag}{2} \sum a(x) b(x).
\end{align}

The massless Hamiltonian has a diagonal shift symmetry for the $b$ fermion \footnote{The translational symmetry in the $i$ th direction is investigated in \cite{Onogi:2025xir}}. We denote it associated with Eq. \eqref{eq: chiral operator} by $T_b$, which acts as
\begin{align}
    T_b a(x)T_b^{-1}=a(x),~
    T_b b(x) T_b^{-1}= (-1)^{x^2} b(x+T).
\end{align}
Other quantized charges are defined as
\begin{align}
    Q_n= T_b^n Q_0 T_b^{-n}= \frac{\imag}{2} \sum f_n(x) a(x)b(x+nT),
\end{align}
where $f_n(x)=(-1)^{nx^2 + \frac{1}{2}n(n-1) }$. Setting
\begin{align}
    G_n=  \frac{\imag}{2} \sum f_n(x) \qty[ a(x)a(x+nT) -b(x)b(x+nT)],
\end{align}
we find a closed algebra under multiplication: 
\begin{align}
    [Q_n, Q_m]=\imag G_{m-n},~[Q_n,G_m]=2\imag(Q_{n-m}-  Q_{n+m}),~[G_n,G_m]=0,
\end{align}
which is the Onsager algebra \footnote{ $Q_0$ and $Q_1$ are equivalent to $Q_V$ and $Q_A$ defined in \cite{Catterall:2025vrx}.}. The unquantized charge of the Onsager algebra is defined as
\begin{align}
    \tilde{Q}= &\frac{1}{2} (Q_1 + Q_{-1} )= \sum_x \chi^\dagger (x) \frac{\Gamma- \Gamma^\dagger }{2 \imag} \chi(x)=\sum_x  \tilde{j}^0(x) \label{eq: Onsager charge}. \\
    \tilde{j}^0(x)=&\frac{1}{2\imag} (\chi^\dagger (x )\Gamma \chi(x)-h.c)=\frac{1}{2} (-1)^{x^2} (\chi^\dagger(x) \chi(x+T)+  \chi^\dagger(x+T) \chi(x) ) \label{eq: Onsager charge density}.
\end{align}

We reveal the physical meaning of $Q_1$. $Q_1$ acts on $a$ and $b$ as
\begin{align}
    [Q_1, a(x)]&= -\imag (-1)^{x^2}b(x+T), \\
    [Q_1, b(x)]&= -\imag (-1)^{x^2}a(x-T).
\end{align}
These equations lead to
\begin{align}
    [Q_1, \psi(r)] \to -\imag (\beta \alpha_2\otimes {}^t \sigma_2) \psi^\ast(r)
\end{align}
in the continuum limit \cite{Onogi:2025xir}. This transformation looks like the charge conjugation \eqref{eq: charge conjugation} with a flavor rotation rather than the chiral transformation. The derivation is implemented in Appendix \ref{sec: taste basis}.

We consider the relation between this algebra and the chiral charge \eqref{eq: chiral charge}. Substituting \eqref{eq: majorana rep} into \eqref{eq: chiral charge}, we get
\begin{align}
    Q_A= \frac{\imag}{4} \sum (-1)^{x^2} (a(x) a(x+T) +b(x )b(x+T) ) \neq  \tilde{Q}%+\frac{\imag}{2} \frac{Q_{1}+ Q_{-1} }{2} .
\end{align}
This charge commutes with all $Q_n,~G_m$, so it can be interpreted as a central charge of the Onsager algebra. Our numerical results shown in the next section indicate that this charge has the chiral anomaly. Note that 
\begin{align}
    \chi^\dagger \Gamma \chi(x) = j_A^0(x)  + \imag \tilde{j}^0(x),\\
    \sum_x \chi^\dagger(x) \Gamma \chi(x)= Q_A+ \imag \tilde{Q}.
\end{align}

\fi

%The first term is Hermitian, and the second one is anti-Hermitian. Note that the Hermitian part commutes with all $Q_n,~G_m$, so it can be interpreted as a central charge of the Onsager algebra. Our numerical result indicates that the chiral anomaly is captured by the Hermitian part rather than the anti-Hermitian part. This result supports the statement in \cite{Onogi:2025xir} that $Q_1$ is not identified as a chiral charge. %Unlike the $(1+1)$-dimensional case, it is not written by generators of the Onsager algebra. 

\section{Interaction with Electromagnetic Field}
\label{sec: Anomaly Eq}
In this section, we consider the chiral anomaly in the presence of a $U(1)$ gauge field. We begin by reviewing the continuum theory and summarizing the results relevant to chiral symmetry in Sec.~\ref{sec: Continuum Theory brief}, while the detailed calculations are presented in Appendix~\ref{app: 3D continuum space}. In Secs.~\ref{sec: lattice theory} and \ref{sec: numerical result}, we investigate the chiral anomaly in the KS Hamiltonian with $U(1)$ link variables.   

\subsection{Continuum theory}
\label{sec: Continuum Theory brief}
We briefly review the single Dirac fermion system with a background $U(1)$ gauge field on the continuum theory \cite{Dunne:1989gp},
\begin{align}
    H=  \int_{\mathbb{R}^3} d^3x~\psi^\dagger h \psi(x,t), ~(h=  \alpha^i \qty(- \imag \partial_i + A_i(x,t))),
\end{align}
where $\alpha^i=\sigma_3\otimes \sigma_i$. We assume that the electric field $E_i= \partial_0 A_i$ and magnetic field $B_i=\epsilon^{ijk} \partial_j A_k $ are constant for $x$ and $t$. 

This system has a chiral symmetry generated by the chiral operator,
\begin{align}
    \gamma_5= -\imag \alpha^1 \alpha^2 \alpha^3=\sigma_3 \otimes 1.
\end{align}
Its charge density and current are given by
\begin{align}
    j_{A,c}^0(x,t)= &\lim_{y\to x} \psi^\dagger (x,t)\gamma^5 e^{\imag \int^x_y A_i(z,t) dz^i } \psi(y,t), \\
    j_{A,c}^i(x,t)=& \lim_{y\to x} \psi^\dagger (x,t) \alpha^i \gamma^5 e^{\imag \int_y^x A_i(z,t) dz^i } \psi(y,t),
\end{align}
where the subscript $c$ means ``continuum''.

We solve the eigenvalue problem of $h$ at fixed time $t$ and find the positive and negative energy states as
\begin{align}
    hu(\Omega, x,t)&=E(\Omega,t)u(\Omega, x,t),~ \\
    hv(\Omega,x,t)&=-E(\Omega,t)v(\Omega,x,t)
\end{align}
for energy $E(\Omega,t)> 0$ \footnote{There is an ambiguity as to whether the zero modes should be regarded as positive or negative energy states.}.  $\Omega$ is a set of all parameters that characterize the wave functions $u$ and $v$. The normalization of them is determined by
\begin{align}
\int d^3x\,
  u^\dagger(\Omega,x,t)\,u(\Omega',x,t)
&= \delta_{\Omega\Omega'}, 
\\
\int d^3x\,
  v^\dagger(\Omega,x,t)\,v(\Omega',x,t)
&= \delta_{\Omega\Omega'}, 
\\
\int d^3x\,
  u^\dagger(\Omega,x,t)\,v(\Omega',x,t)
&= 0,
\\
\int d\Omega\,
\Bigl[
  u(\Omega,x,t)\,u^\dagger(\Omega,x',t)\nonumber
 \\+ v(\Omega,x,t)\,v^\dagger(\Omega,x',t)
\Bigr]
&= \delta^{(3)}(x-x').
\end{align}

 %If zero modes exist, it is an ambiguity as to which ones are made negative energy. In this paper, we assume that zero modes with $C=+1$ are negative energy states.

Let us expand the Dirac fermion field by the creation and annihilation operators of the wave functions:
\begin{align}
    \psi(x,t)=  \int d\Omega \qty[ u(\Omega,x,t) b(\Omega,t) + v(\Omega,x,t) d^\dagger(\Omega,t) ],
\end{align}
where the annihilation operators $b(\Omega)$ and $d(\Omega)$ satisfy the standard anticommutation relations:
\begin{align}
    \qty{ b(\Omega,t), b^\dagger(\Omega^\prime,t)  }= \qty{ d(\Omega,t), d^\dagger(\Omega^\prime,t)  }=\delta_{\Omega\Omega^\prime} 
\end{align}
and the others are zero. The vacuum state at time $t$ is defined as
\begin{align}
    b(\Omega,t) \ket{0,t}=d(\Omega,t) \ket{0,t}=0. 
\end{align}

\fix{According to the argument in Appendix \ref{app: 3D continuum space}}, the vacuum expectation value of $j_A^0$ is given by
\begin{align}
     \ev{j_{A,c}^0(x,t)}:=&\bra{0,t} j_{A,c}^0(x,t) \ket{0,t} \nonumber\\
     =&\lim_{y\to x}  \int d\Omega v^\dagger (\Omega, x) \gamma^5 v (\Omega, y) \nonumber \\
     =&\lim_{y\to x} \imag  \frac{B_i }{ 2\pi^2}  \frac{(y-x)^i}{ \norm{ y-x }^2 } \label{eq: 3D ev cont}.
\end{align}
As a result of the point splitting, the value becomes a pure imaginary number. Furthermore, the chiral current satisfies the chiral anomaly equation,
\begin{align}
    \ev { \partial_\mu j_{A,c}^ \mu(x,t)}= - \frac{E_i B_i }{2\pi^2} ,
\end{align}
and
\begin{align}
    \dv{}{t} \ev{Q_{A,c}(t)}= -\sum_x  \frac{E_i B_i }{2\pi^2} \label{eq: 3D anomaly eq cont}.
\end{align}

\subsection{Lattice theory}
\label{sec: lattice theory}

In the presence of link variables, the chiral operator \eqref{eq: chiral operator} does not commute with the massless Hamiltonian \footnote{If we consider the dynamical link variables, there is a shift symmetry under $\chi(x) \to \imag (-1)^{x^2}\chi(x+T)$ and $U_i(x)\to U_i(x+T)$ \cite{Golterman:2024xos}. }. However, there is a certain magnetic field configuration in which $\Gamma$ commutes. Imposing the electric field adiabatically, we solve the eigenvalue problem of the Hamiltonian and determine the vacuum state. Then, we show that the expectation value of the chiral charge density \eqref{eq: chiral charge} satisfies the anomaly equation in the kinetic normal ordering \cite{Dunne:1989gp}.

%\subsection{$(3+1)$-dimension}
%We proceed to the $(3+1)$-dimensional KS fermion system. Here, we put this system in a uniform background magnetic field and investigate the response of the chiral charge to an electric field. We show that the expectation value of the chiral charge satisfies a chiral anomaly equation \cite{Dunne:1989gp} on the lattice space. 

%In the continuum theory, the chiral charge evolves with time according to an anomaly equation \cite{Dunne:1989gp}. 

We fix link variables as
\begin{align}
    U_1(x,t)&= \begin{cases}
    e^{\imag B x^3} & ( x^1 \neq N-1 ) \\
    e^{\imag B x^3}e^{-\imag B N x^2 } & (x^1=N-1)
    \end{cases} ,~ \\
    U_2(x,t)&= e^{\imag B (x^1-x^3)},~ \\
    U_3(x,t)&=\begin{cases}
    e^{ \imag \frac{2\pi}{N} t} & ( x^3 \neq N-1 ) \\
    e^{\imag B N ( x^1-x^2)} e^{\imag \frac{2\pi}{N} t} & (x^3=N-1)
    \end{cases} ,
\end{align}
where $B$ is a magnetic flux through one plaquette and takes a discrete value
\begin{align}
    B=\frac{2\pi n}{N^2}~(n\in \mathbb{Z})
\end{align}
from the periodic boundary condition for the KS fermion. This configuration generates a uniform magnetic field $B$ and electric field $E$:
\begin{align}
    B_1=B_2=B_3=B,~E_1=E_2=0,~E_3= \frac{2\pi}{N}.
\end{align}
In the presence of the link variables, the shift operators are modified as
\begin{align}
    T^U_i \chi(x,t)= \eta_i(x) U_i(x,t) \chi(x+\hat{i},t). 
\end{align}
Since $T_i^U T_j^U + T_j^U T_i^U\neq 2 \delta_{ij}(T_j^U)^2$ in general, the existence of the chiral symmetry is nontrivial. However,
\begin{align}
    [T_i^U, T_1^UT_2^UT_3^U]=0
\end{align}
holds under our link variables. Then, we can define the chiral transformation commuting with the massless Hamiltonian in the presence of the magnetic field as
\begin{align}
    \Gamma^U \chi(x,t)=(-\imag) e^{-\imag B/2} T_1^U T_2^U T_3^U \chi(x,t) \label{eq: chiral operator with mag}.%=\imag(-1)^{x^2} e^{-\imag B/2}T_1 T_2 T_3 \chi(x,t) 
\end{align}
Here, $e^{-\imag B/2}$ is the normalization factor so that 
\begin{align}
    (\Gamma^U)^N= (-1)^n e^{\imag 2\pi t}.
\end{align}
Note that the KS fermion effectively becomes an antiperiodic function when $n$ is odd. We also define the chiral charge $Q_A$ and its density $j_A^0(x,t)$ as well as Eq. \eqref{eq: chiral charge}. 

\subsection{Numerical results}
\label{sec: numerical result}

\begin{figure*}
\begin{minipage}[b]{0.49\linewidth}
\centering
\includegraphics[width=\linewidth]{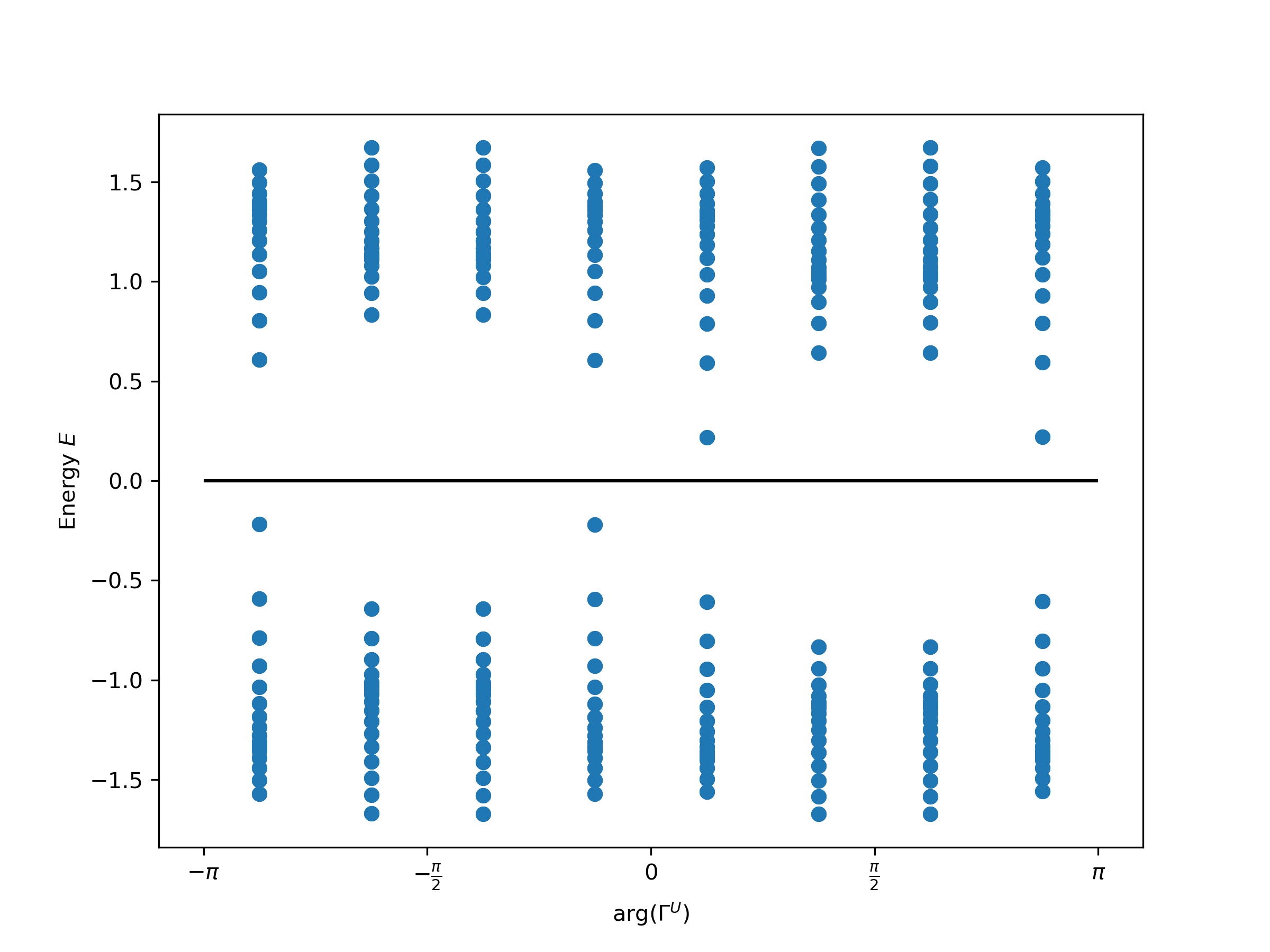}
\end{minipage}
\hfill
\begin{minipage}[b]{0.49\linewidth}
    \centering
    \includegraphics[width=\linewidth]{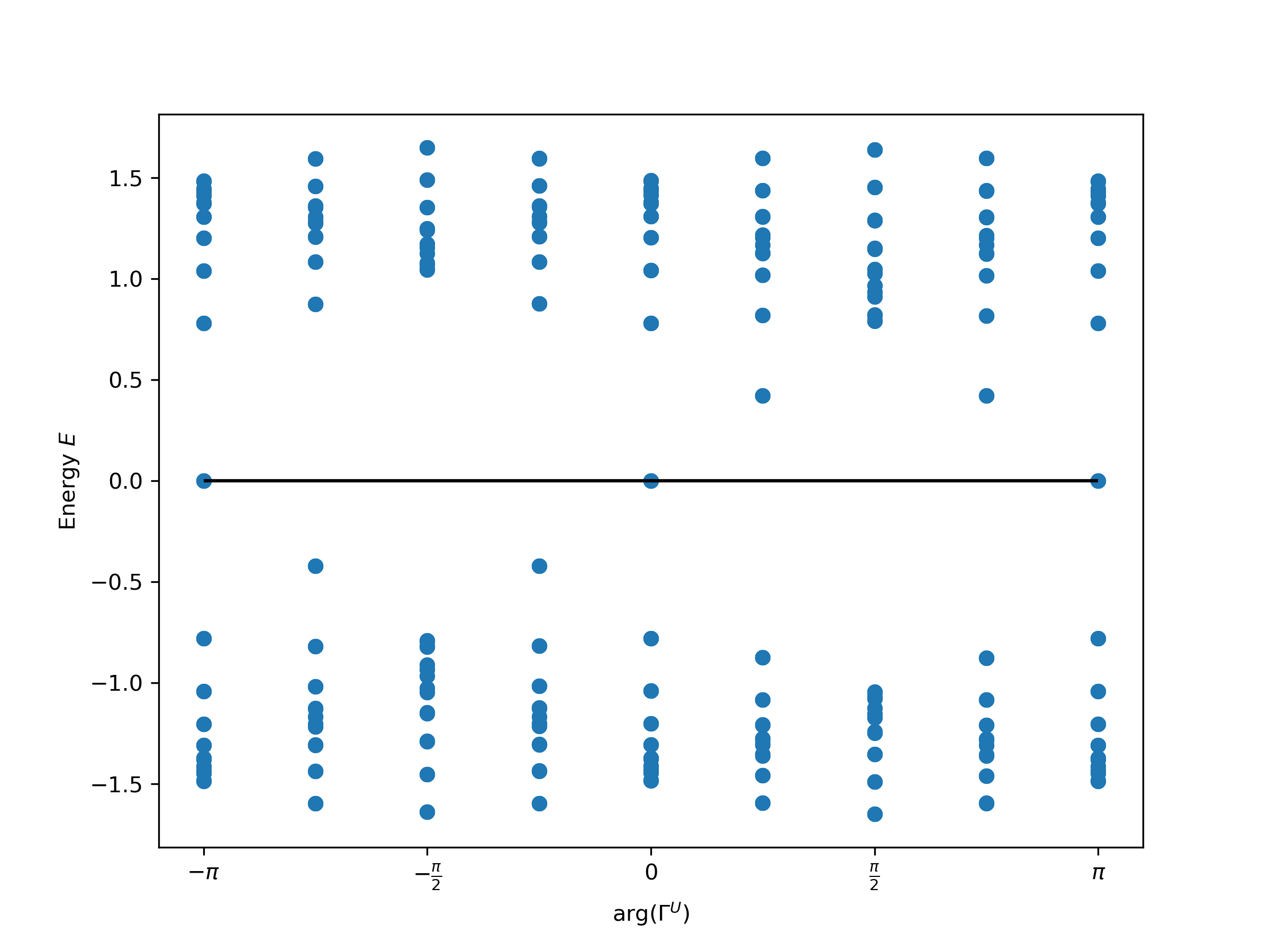}
    \end{minipage}
    \caption{Energy spectrum represented by the argument (phase) of $\Gamma^U$ at $N=8$, $m=0$, and $t=0$. The left and right panels correspond to $n = -1$ and $n = 2$, respectively.}
    \label{fig: energy vs chirality}
\end{figure*}

We solve the eigenvalue problem of the one-particle Hamiltonian $h$ with $m=0$ on each time slice. Unlike the case $(1+1)$-dimensional, it is hard to solve the eigenvalue problem of the Hamiltonian analytically, so we only calculate it numerically. Fixing $N=8,~m=0$ and $t=0$, the energy is plotted against the eigenvalue of $\Gamma^U$ in Fig.~\ref{fig: energy vs chirality} with $n=-1$ and $ n=2$. When $n=-1$, we can see the gap generated by the antiperiodicity $(\Gamma^U) ^N=-1$. On the other hand, when $n=2$, there are zero modes at $\Gamma^U=+1$ and $\Gamma^U= e^{\pm \imag \pi}=-1$. 

\begin{figure*}
\begin{minipage}[b]{0.49\linewidth}
\centering
\includegraphics[width=\linewidth]{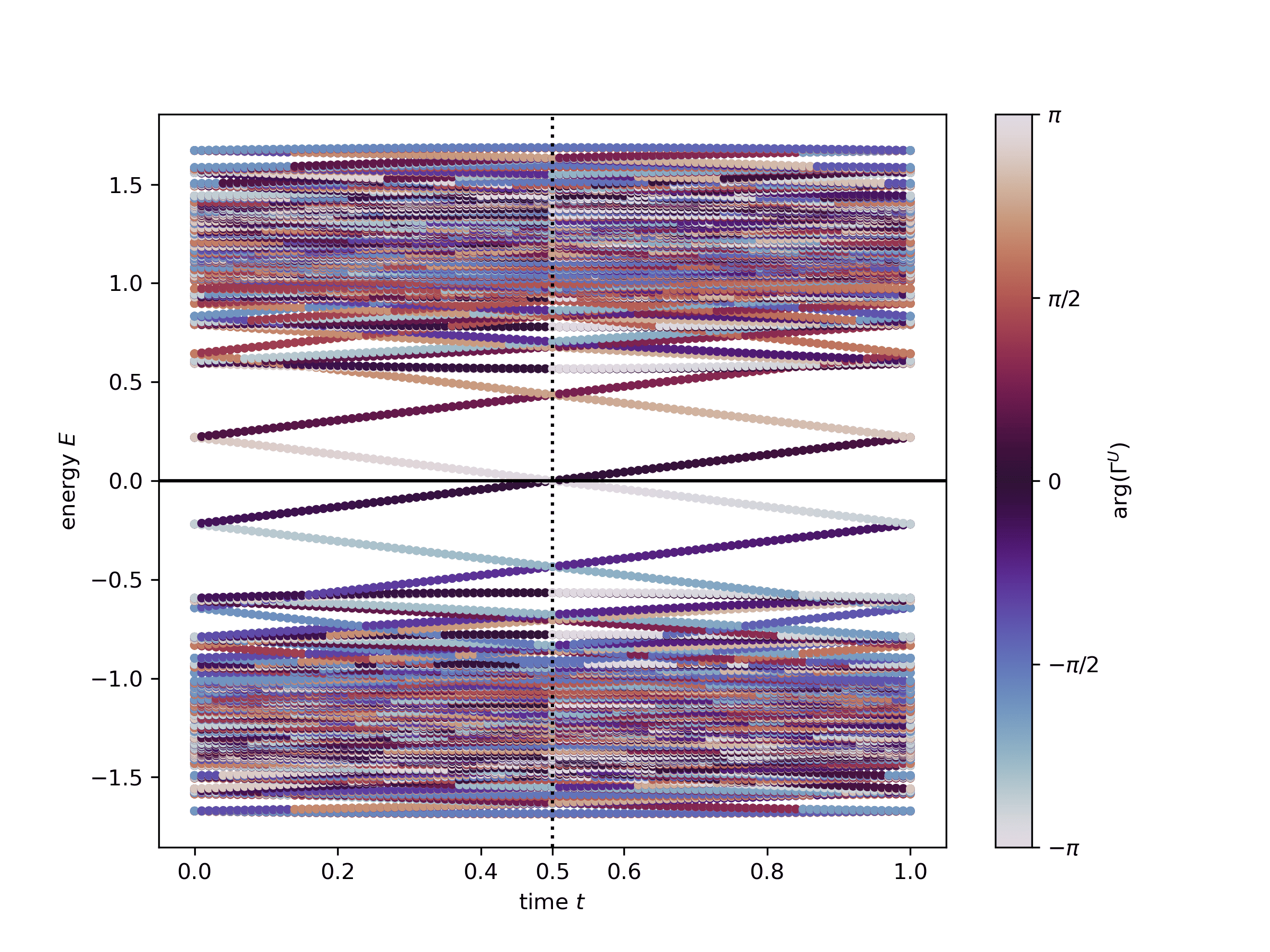}
\end{minipage}
\hfill
\begin{minipage}[b]{0.49\linewidth}
    \centering
    \includegraphics[width=\linewidth]{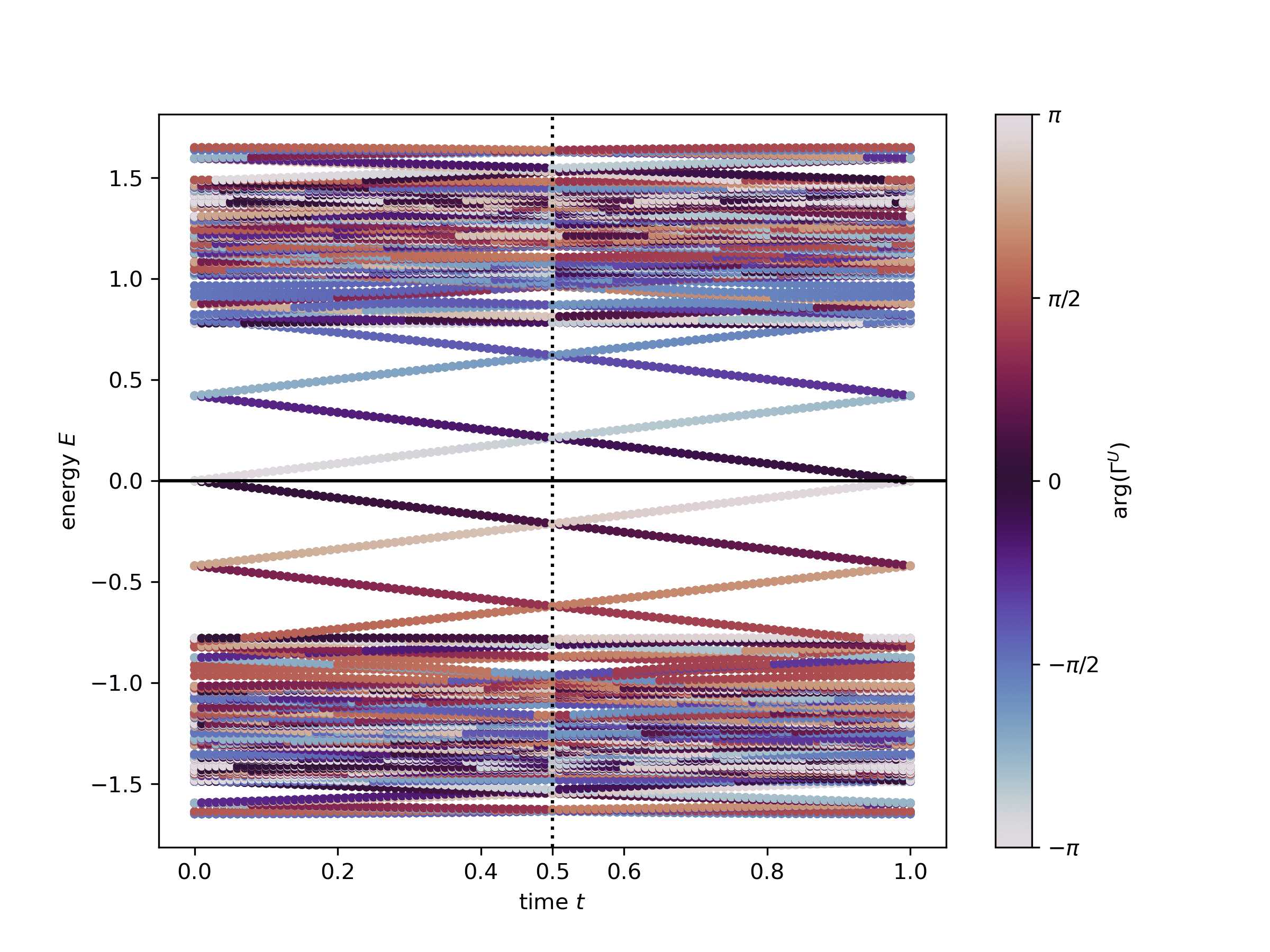}
    \end{minipage}
\caption{Time evolution of the energy spectrum at $N=8$ and $m=0$. The left and right panels correspond to $n = -1$ and $n = 2$, respectively. The color gradation represents the eigenvalues of $\Gamma^U$: the positive and negative imaginary parts are shown in red and blue, respectively, while the real part is encoded in the brightness.}
    \label{fig: energy vs time}
\end{figure*}

We also investigate the time evolution of the energy, shown in Fig.~\ref{fig: energy vs time}, with $N=8$ and $m=0$. The color gradation represents the eigenvalue of $\Gamma^U$: a positive imaginary part is indicated in red, a negative one in blue, and the real part is encoded in the brightness. We observe that the $\Gamma^U=\pm 1$ modes cross $E=0$ at $t=0.5$ for $n=-1$, and at $t=0$ and $t=1$ for $n=2$. Our numerical results confirm that zero modes appear at $t\in \mathbb{Z}+\frac{n}{2}$.\footnote{We do not provide an analytic proof of this statement.}

\if
Let $u$ and $v$ be positive and negative energy wave functions such that
\begin{align}
    hu(\Omega, x,t)=E(\Omega)u(\Omega, x,t),~ hv(\Omega,x,t)=-E(\Omega)v(\Omega,x,t)
\end{align}
for energy $E(\Omega)> 0$. $\Omega$ is a set of all parameters that characterize the wave functions $u$ and $v$. The normalization of them is determined by
\begin{align}
    \sum_x~ u^\dagger (\Omega,x,t)  u(\Omega^\prime,x,t)=\sum_x~ v^\dagger (\Omega,x,t)  v(\Omega^\prime,x,t)= \delta_{\Omega\Omega^\prime}, \\
    \sum_x~ u^\dagger (\Omega,x,t)  v(\Omega^\prime,x,t)=0, \\
    \sum_{\Omega} \qty[  u(\Omega,x,t) u^\dagger (\Omega,x^\prime,t) +v(\Omega,x,t) v^\dagger (\Omega,x^\prime,t)   ] = \delta_{xx^\prime}.
\end{align}
 If zero modes exist, it is an ambiguity as to which ones are made negative energy. In this paper, we assume that zero modes with $C=+1$ are negative energy states.

Let us expand the KS fermion field by the creation and annihilation operators of the wave functions:
\begin{align}
    \chi(x,t)=  \sum_{ \Omega } \qty[ u(\Omega,x,t) b(\Omega,t) + v(\Omega,x,t) d^\dagger(\Omega,t) ],
\end{align}
where the annihilation operators $b(\Omega)$ and $d(\Omega)$ satisfy the standard anticommutation relations:
\begin{align}
    \qty{ b(\Omega,t), b^\dagger(\Omega^\prime,t)  }= \qty{ d(\Omega,t), d^\dagger(\Omega^\prime,t)  }=\delta_{\Omega\Omega^\prime} 
\end{align}
and the others are zero. The vacuum state at time $t$ is defined as
\begin{align}
    b(\Omega,t) \ket{0,t}=d(\Omega,t) \ket{0,t}=0.
\end{align}

The vacuum expectation value of the chiral charge density is given by
\begin{align}
    \ev{j_A^0(x,t)}:=\bra{0,t} j_A^0(x,t) \ket{0,t}= \sum_{\Omega} v^\dagger(\Omega,x,t) C v (\Omega,x,t) 
\end{align}
\fi
%This function is smooth when $m\neq 0$. However, in the massless limit, its value jumps when $t$ is an integer. From the properties of the eigenstates, $\ev{j_5^0(x,t)}$ is a periodic function with $t+1 \sim t$, and it is sufficient to assume $0\leq t <1$.

\begin{figure*}
\begin{minipage}{0.49\linewidth}
\centering
\includegraphics[width=\linewidth]{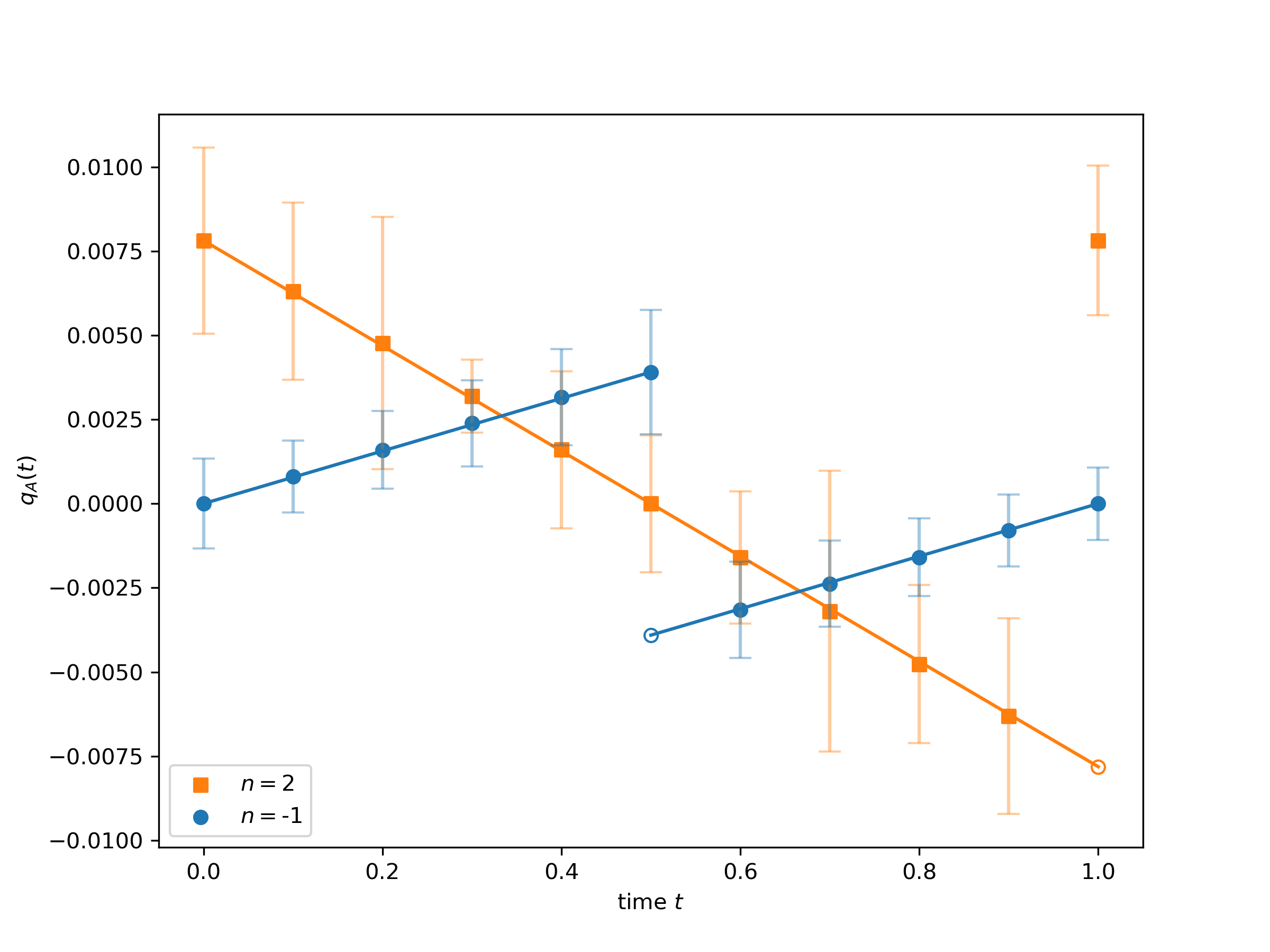}
\end{minipage}
\hfill
\begin{minipage}{0.49\linewidth}
    \centering
    \includegraphics[width=\linewidth]{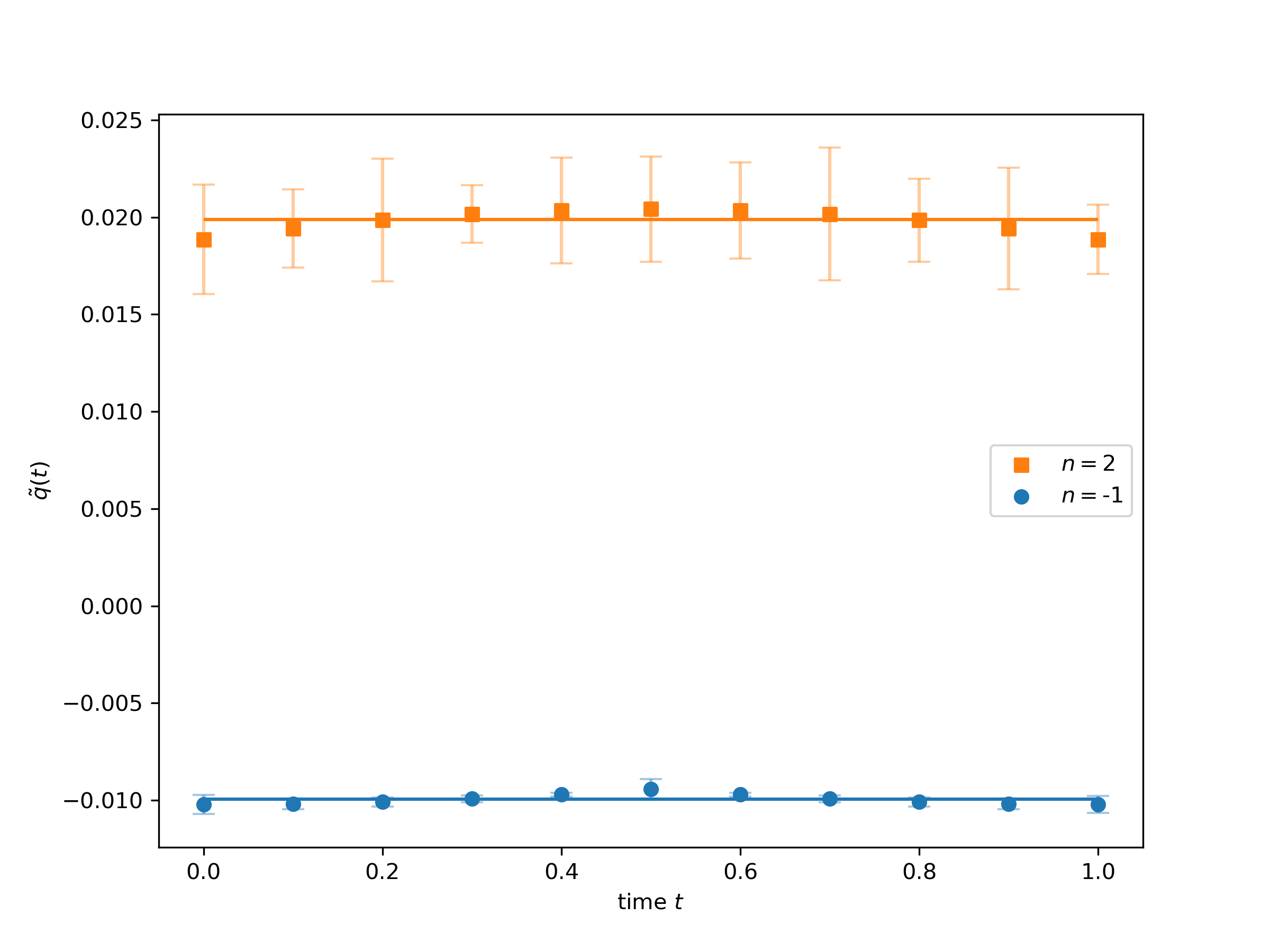}
    \end{minipage}
\caption{Left panel: time evolution of the expectation value of the chiral charge density $j_A^0$ under a magnetic field in the diagonal direction and an electric field applied along the $x^3$-axis. Filled symbols indicate spatially averaged values, with error bars showing spatial variations. Solid lines represent the fitting functions defined in Eq.~\eqref{eq: fitting function qA}. The white circle marks the discontinuity. Right panel: the same plot for $\tilde{j}^0$, where the fitting function is defined in Eq.~\eqref{eq: fitting function tq}. We fix $N=8$ and $m=0$, and choose $n=-1$ and $n=2$.
}
    \label{fig: q5_3D}
\end{figure*}

We compute the expectation values of $j_A^0$ and $\tilde{j}^0$ in the same manner as in Sec.~\ref{sec: Continuum Theory brief}. Here, the zero modes with $\Gamma^U = 1$ are assigned to the negative energy states to define the vacuum. The spatially averaged values are plotted as functions of time in the left and right panels of Fig.~\ref{fig: q5_3D}, where the parameters are fixed in the same way as in Fig.~\ref{fig: energy vs chirality}. The averages are defined by
\begin{align}
    q_A(t)=& \frac{1}{N^3}\sum_x \ev{j^0_A(x,t)}= \frac{1}{N^3} \ev{Q_A(t)} ,~ \\
    \tilde{q}(t)=& \frac{1}{N^3}\sum_x \ev{\tilde{j}^0(x,t)}= \frac{1}{N^3} \ev{\tilde{Q}(t)} \label{eq: average},
\end{align}
%whose real parts are expressed by orange dots and imaginary parts are depicted by blue dots. 
and error bars are given by
\begin{align}
    \sigma_A(t)&= \sqrt{ \frac{1}{N^3}\sum_x[ \ev{j_A^0(x,t)} -q_A (t)] ^2 } ,~\\
    \tilde{\sigma}(t)&= \sqrt{ \frac{1}{N^3}\sum_x[ \ev{\tilde{j}^0(x,t)} -\tilde{q}(t) ] ^2 }. \label{eq: sd}
\end{align}
%If $\sigma$ is not equal to zero, $\ev{j_5^0(x,t)}$ changes with $x$. %The two solid black lines are the relevant terms of Eq. \eqref{eq: analytic calculation}. For the most part, the numerical result coincides with the analytic result, but there are some deviations at $t=0,1$ due to the existence of zero modes.
%The definition of the average and error bars are the formulas \eqref{eq: average} and \eqref{eq: sd} with $N$ replaced by $N^3$. This result implies 
The solid lines represent the fitting functions defined as
\begin{align}
    %\imag  \frac{2B}{2 \pi^2} - \frac{2 \abs{B} }{2 \pi^2}E_3 \qty    {\frac{n}{\abs{n}} t - \frac{n}{2} } + \order{ \frac{1}{N^4} }   \\ 
    q_{A,\text{fitting}}(t)&=%\imag  \frac{2n}{\pi N^2 } 
    -  \frac{4 \abs{n} }{ N^3 } \qty(\qty{ \frac{n}{\abs{n}} t -\frac{n}{2}} -\frac{1}{2}), \label{eq: fitting function qA}\\
    \tilde{q}_\text{fitting}(t)&=\frac{2n}{\pi N^2 } \label{eq: fitting function tq},
\end{align}
where the curly bracket is the sawtooth function, which takes the fractional part of the given real number. We find that our lattice data are in good agreement with Eq.~\eqref{eq: fitting function qA}. The $q_A(t)$ exhibit discontinuities at $t \in \mathbb{Z} + \tfrac{n}{2}$, where zero modes appear. This phenomenon is consistent with the interpretation in \cite{Isler:1987ax,Manton:1985jm,Nielsen:1991si,Ambjorn:1983hp}. The electric field accelerates particles, leading to the emergence of positive energy excitations from the Dirac sea, while others are driven into negative energy states and sink back into the sea. As a result, the chirality of the Dirac sea decreases discontinuously. Note that the value of $q_A(t)$ at $t\in \mathbb{Z}+\frac{n}{2}$ depends on the choice of the definition of positive and negative energy states.

%We can see that the real parts jump several times. Since the time at which zero modes appear depends on whether $n$ is even or odd, the jump time also changes.   

Our numerical results are in good agreement with the continuum predictions. The time derivative of $\ev{Q_A(t)}$ in the continuous region is estimated as
\begin{align}
    \dv{}{t} \ev{Q_A(t)}\simeq  -  \sum_x \frac{4 n }{ N^3 }= - 2\sum_x \frac{B_i E_i }{ 2\pi^2}.
\end{align}
Compared with the single Dirac fermion case \eqref{eq: 3D anomaly eq cont}, this equation is equivalent to the chiral anomaly equation of a two-flavor Dirac fermion. %Thus, $Q_A(t)$ in Eq.~\eqref{eq: chiral charge} can be interpreted as the lattice version of the chiral charge. 

On the other hand, what does $\tilde{Q}$ or $\tilde{j}^0$ mean? The expectation value is estimated as
\begin{align}
    \ev{\tilde{j}^0(x,t) } \simeq \tilde{q}_{\text{fitting}}= 2\frac{B}{ 2\pi^2},
\end{align}
which is equivalent to the imaginary part of Eq.~\eqref{eq: 3D ev cont} when $y=x+T$ and $B_i=B$. The factor $2$ means the number of Dirac fermions in the continuum limit. Thus, we conclude that 
\begin{align}
\ev{ j^0_\text{reg}(x,t)} &= \ev{j_A^0(x,t)}+ \imag \ev{\tilde{j}^0(x,t)} \to 2 \ev{j_{A,c}^0(x,t)}, \\
\dv{}{t} \ev{ Q_\text{reg} (t)} &= \dv{}{t} \qty(  \ev{Q_A(t)} + \imag \ev{\tilde{Q}(t)}) \to 2  \dv{}{t} \ev{Q_{A,c}(t)}
\end{align}
and the diagonal shift operator $\Gamma^U$ converges to the chiral operator $\gamma_5\otimes 1$ in the continuum limit.

\if0
We compare our numerical result with the continuum prediction derived in Appendix \ref{app: 3D continuum space}. In the continuum single Dirac fermion system, the chiral current is defined by
\begin{align}
    j_A^0(x)= \lim_{y\to x} \psi^\dagger(x) \gamma^5 \psi(y) .
\end{align}
Its expectation value is given by \cite{Dunne:1989gp}
\begin{align}
    \ev{j_A^0(x)}= \lim_{y\to x} \imag  \frac{B_i}{ 2\pi^2}     \frac{(y-x)^i }{ \norm{ y-x }^2 } 
\end{align}
Now, the point splitting is clearly determined as $y=x+T$. 
Substituting $y=x+T$ into Eq.~\eqref{eq: 3D ev cont}, we have
\begin{align}
    \ev{j_A^0(x)}=  \imag  \frac{3B}{ 2\pi^2} \frac{ 1}{ 3 }=\imag  \frac{B}{ 2\pi^2}= \imag \frac{n}{\pi N^2} .
\end{align}
Compared with it, the imaginary part of our numerical result is off by a factor of two. Since the KS fermion system corresponds to the two-flavor Dirac fermion system, we conclude that our imaginary part is consistent with the continuum prediction. Moreover, the real part of the chiral charge implies that the chiral anomaly equation \cite{Dunne:1989gp} on two-flavor Dirac fermion
\begin{align}
    \dv{}{t} \ev{Q_A(t)} = - 4n = - 2\sum_x \frac{B_i E_i }{ 2\pi^2}.
\end{align}
holds \footnote{The definition of a magnetic field in \cite{Dunne:1989gp} differs from our definition only in sign.}.
\fi 

One may be wondering whether our definition can capture the chiral anomaly under link variables that break the commutativity with $H$ and $Q_A$. We try the same computation under 
\begin{align}
    U_1(x,t)=1,~U_2(x,t)=e^{\imag B x^1 },~U_3(x,t)=e^{\imag \frac{2\pi}{N}t},
\end{align}
which generate a uniform magnetic and electric fields in the $x^3$ direction. We can define the chiral operator \eqref{eq: chiral operator with mag}, but it no longer commutes with the Hamiltonian. 

\begin{figure*}
\begin{minipage}[b]{0.49\linewidth}
\centering
\includegraphics[width=\linewidth]{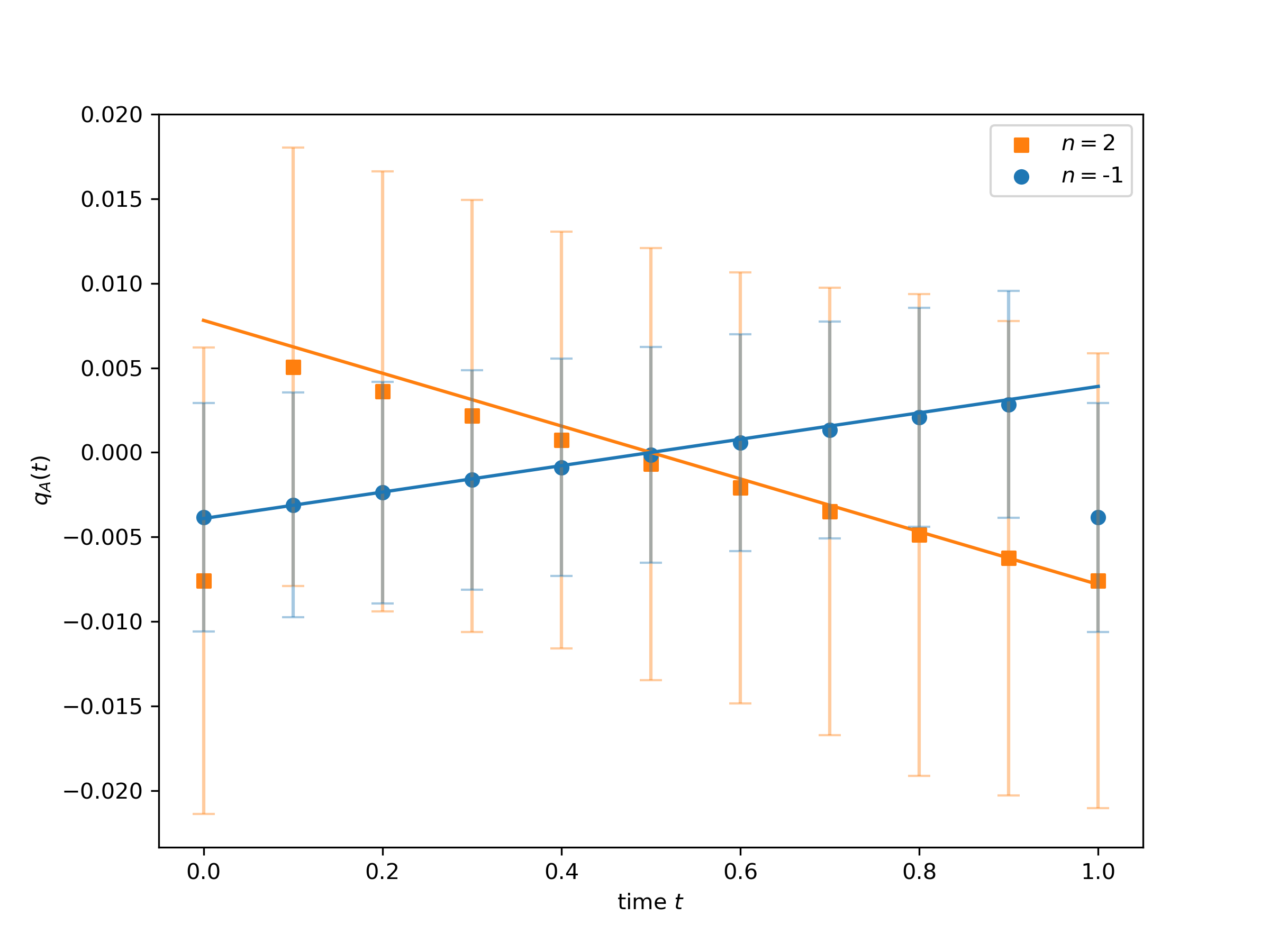}
\end{minipage}
\hfill
\begin{minipage}[b]{0.49\linewidth}
    \centering
    \includegraphics[width=\linewidth]{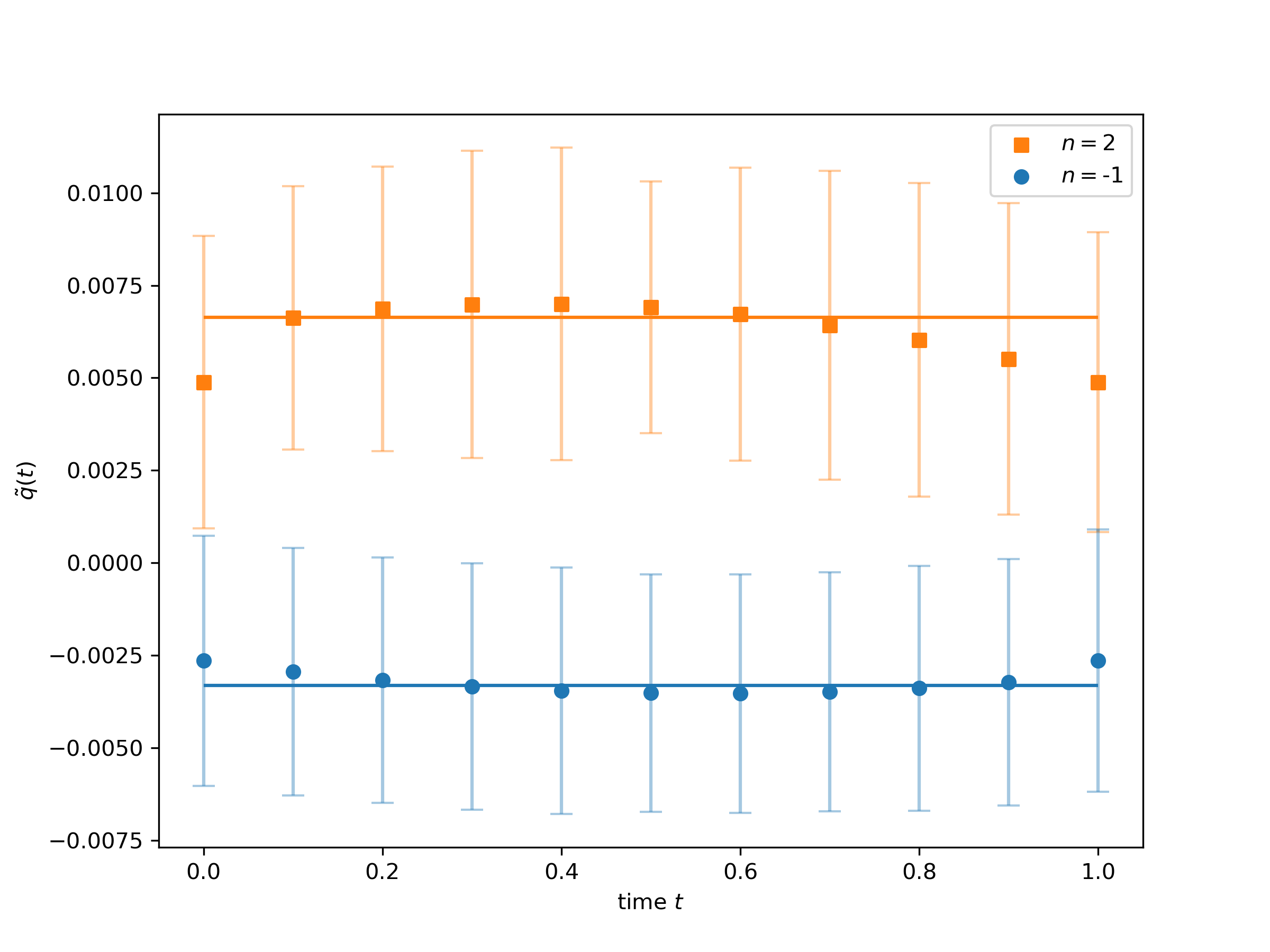}
    \end{minipage}
\caption{Same plots as in Fig.~\ref{fig: q5_3D}, but with both the magnetic and electric fields aligned along the $x^3$-axis.}
    \label{fig: q5_B3_3D}
\end{figure*}

Anyway, we plot the expectation value of the chiral charge density at $N=8, n=-1$ and $n=2$ in Fig.~\ref{fig: q5_B3_3D}. The deviations are bigger than the previous examples. This reflects the fact that the chiral operator is not an exact symmetry of the Hamiltonian. Nevertheless, the averages agree with the continuum predictions indicated by solid lines:
\begin{align}
    q_{A,\text{fitting}}(t) &=  -  2 \frac{B_3E_3}{2\pi^2} t \quad ( 0<t<1 ), \\
    \tilde{q}_{\text{fitting}}(t)&= 2 \frac{B_3}{ 2\pi^2} \frac{1}{ 3 }.
\end{align}
Thus, these results agree with the continuum prediction.
%This result supports our conclusion. %This implies that the $j_A^0(x,t)$ can be interpreted as the chiral charge density on the lattice space.

\section{Conclusion}
\label{sec: Conclusion}
We studied the $(3+1)$-dimensional KS fermion system in Hamiltonian formalism. We confirmed that the diagonal shift operator can be interpreted as the chiral operator on the lattice.

%我々は$(3+1)$次元のKS fermionの系を解析し、Hamiltonianを考えた。この系では立方体の対角線方向への並進が$U_A(1)$変換であると期待されていた。我々は数値計算によって、この並進演算子がカイラル演算子と見做せることを確かめた。

%In the Sec.~\ref{sec: Anomaly Eq}, 
In the presence of link variables, the diagonal shift symmetry is violated because of their spatial dependence. Nevertheless, we found a specific configuration that preserves the diagonal shift symmetry in Sec.~\ref{sec: Anomaly Eq}. In this case, the magnetic field aligns with the diagonal direction, consistent with the point-splitting construction of the chiral charge on the continuum space. Note that the diagonal shift operator is unitary rather than Hermitian. We defined two charges $Q_A$ and $\tilde{Q}$ corresponding to the Hermitian and anti-Hermitian parts of the diagonal shift operator. We numerically calculated the expectation values of $Q_A$ and $\tilde{Q}$ and confirmed two things. One is that the time evolution of $\ev{Q_A}$ satisfies the chiral anomaly equation for a two-flavor Dirac fermion. The other is that the expectation value of $\tilde{Q}$ is consistent with the continuum prediction. Thus, we concluded that the diagonal shift operator can be interpreted as the lattice regularized chiral charge. This implies that $Q_A$ is equivalent to the Chern–Simons term \cite{Tatsumi:1999jlf, Bietenholz:2000ca} cohomologically.

%The imaginary part agrees with the continuum prediction, where it originates from the UV regularization. We further introduce an adiabatic electric field and investigate the time evolution of the chiral charge. Our numerical results show that only the real part oscillates in a sawtooth pattern, and the chiral anomaly equation for two-flavor Dirac fermions is satisfied. This implies that $Q_A$ corresponds to the Chern–Simons term \cite{Tatsumi:1999jlf, Bietenholz:2000ca} in the cohomology group.

%Sec.~\ref{sec: Anomaly Eq} で指摘したように、このような並進演算子はlink変数の存在化では、link変数の座標依存性によって、一般にHamiltonianと可換ではない。しかしながら、我々はこのような並進演算子がHamiltonianと可換になるようなlink変数の配位を見つけた。偶然にも、磁場は対角線方向を向いており、point splittingと一致していた。chiral 変換は格子上ではHermitianよりもむしろunitaryであるため、chiral charge densityの期待値は複素数となる。一方で、連続理論においても、無限大の正則化の過程で、Hermitian なカイラル演算子の期待値は複素数になり得るが、我々の数値計算の結果は連続理論の予測とよく一致していた。また、その実部は時間に依存したsawtooth関数で近似されるが、chiral anomaly equationを満たしていた。つまりは実部の期待値はChern-Simons term \fix{reference}とcohomology equivalentであることを示唆している。

We also investigated the relation with the Onsager algebra. In Sec.~\ref{sec: Onsager Algebra}, we constructed the Onsager algebra associated with all shift operators in the three-dimensional space and found that $Q_A$ and $\tilde{Q}$ belong to the Onsager algebra as $\tilde{Q}=(Q_{S_T}+Q_{S_T^{-1}})/2$ and $Q_A=G_{S_T,-S_T}/2$. Since $Q_A$ commutes with all elements in the algebra, it can be interpreted as a central charge. This result supports the statement in \cite{Onogi:2025xir} that $Q_{S_T}$ is not identified as the chiral charge. In the first place, the chiral anomaly is calculated from a triangle diagram with two vector and one axial currents. We need to take into account the current algebra \cite{Johnson:1966se,Adler:1969ccs, Dunne:1989gp}.

%We also construct the Onsager algebra associated with the diagonal shift transformation in Sec.~\ref{sec: Onsager Algebra}. The anti-Hermitian part of our chiral charge is written in $Q_1$ and $Q_{-1}$, but the Hermitian part does not belong to the algebra. Since it commutes with all elements in the algebra, it can be interpreted as a central charge of the algebra. This result supports the statement in \cite{Onogi:2025xir} that $Q_1$ is not identified as the chiral charge. In the first place, the chiral anomaly is calculated from a triangle diagram with two vector and one axial currents. We need to take into account the current algebra \cite{Johnson:1966se,Adler:1969ccs, Dunne:1989gp}.

%加えて、対角方向の並進演算子を使うことでOnsager代数も構成した。我々の定義したchiral chargeの虚部は$Q_1$と$Q_{-1}$の平均として与えられたが、実部はOnsager代数の元で書けなかった。しかしながら、Onsager代数の生成元すべてと可換だったので、中心拡大と解釈可能である。この結果は$Q_1$がaxial chargeではないという主張\cite{Onogi:2025xir}を支持している。そもそも、chiral anomalyは三角diagramから計算されるが、これは一つのaxial $U(1)$ current と二つの$U(1)$ currentの非可換性が重要なので、charge densityだけでなく、currentも考慮すべきである\cite{Johnson:1966se,Adler:1969ccs, Dunne:1989gp}。

%Our result implies that it is possible to construct the discrete chiral symmetry in QED as well as the Schwinger model \cite{Dempsey:2022nys}. 

Our results suggest that the construction of a discrete chiral operator \cite{Dempsey:2022nys} performed in the Schwinger model is also possible in higher dimensions. As in the $(1+1)$ dimension, a shift operator in a specific direction induces a chiral charge in the $(3+1)$ dimension. Therefore, if we clarify the relationship between this chiral charge and the discrete chiral operator, we can expect to construct a quantum electrodynamics Hamiltonian that explicitly possesses chiral symmetry. This will significantly improve the convergence of numerical calculations to the continuum limit, which is likely to contribute to the simulation of physical phenomena involving theta terms.

%3+1次元のmass shift
%我々の結果はSchwinger模型で行われた離散カイラル変換の構成\cite{Dempsey:2022nys}が高次元でも可能であることを示唆する。$(1+1)$次元の場合と同様に、$(3+1)$次元でも特定の方向の並進演算子がchiral 変換を誘導したので、この変換と離散カイラル変換の関係を明らかにすれば、カイラル対称性を顕に備えたQEDのHamiltonianの構成が期待できる。数値計算の連続極限への収束性が大幅に改善し、theta項が関わるような物理現象のシミュレーションに貢献できると思われる。
%$(1+1)$次元のときはmass shiftと呼ばれるカウンター項が離散カイラル変換の対称性を保証したが、これが数値計算の連続極限への収束性を大幅に改善した。もし$(3+1)$次元でも離散カイラル変換による対称性を備えたHamiltonainを構成できたら、

%経路積分形式ではtheta termがあるとき符号問題が起こるが、Hamlitonian形式では起こらないので、theta termが関わる現象の解明につながると思われる。

\section*{Acknowledgment}

We would like to thank S.~Aoki, Y.~Furukawa, H.~Fujii, K.~Fujikura, M.~Honda, Y.~Ikeda, S.~Iso, T.~Numasawa, T.~Okuda, J.~W.~Pedersen, S.~Shimamori, and T.~Tada for valuable and enlightening discussions. S.A. is supported by the RIKEN Special Postdoctoral Researchers Program and JSPS KAKENHI Grant Number 25K17382. T.T. is supported by JST SPRING, Grant Number JPMJSP2108.

\section*{Data availability}
The data that support the findings of this article are not publicly available upon publication because it is not technically feasible and/or the cost of preparing, depositing, and hosting the data would be prohibitive within the terms of this research project. The data are available from the authors upon reasonable request.

\appendix

\section{Kogut-Susskind Hamiltonian on the taste basis}
\label{sec: taste basis}
\setcounter{equation}{0}
The KS fermion field $\chi(x)$ is a single-component fermion. However, KS fermions on $2^3=8$ cube sites are equivalent to $2^{\frac{3-1}{2}}=2$-flavor Dirac fermions, which is the so-called "taste basis" \cite{Burden:1986by, Catterall:2025vrx}. The Dirac fermions live on the blocked lattice whose sites are labeled by $r$, and internal cubic lattice sites are identified by $A$ whose component takes the value $0$ or $1$. Then the original lattice sites are $x=2r+A$. Now, we set $U_i(x) = 1$ and introduce the Dirac fermion fields as 
\begin{align}
    u_{\alpha,f}(r) &= \frac{1}{2} \sum_{A} \chi(2r + A)\, \sigma^{(A)}_{\alpha f}, \\
    d_{\alpha,f}(r) &= \frac{1}{2} \sum_{A} \chi(2r + A)\, \epsilon(A)\, \sigma^{(A)}_{\alpha f},
\end{align}
where $\alpha$ and $f$ denote the spinor and flavor indices, respectively. $u$ and $d$ correspond to the upper and lower components of the $2^{\frac{3+1}{2}}=4$-component Dirac spinor \footnote{When the spatial dimension is $D$, a $2^{\frac{D-1}{2}}$-flavor Dirac fermion with $2^{\frac{D+1}{2}}$ components emerges~\cite{Catterall:2025vrx}.
}. $\sigma^{(A)}$ is a Pauli matrix determined by  
\begin{align}
    \sigma^{(A)}= \sigma_1^{A_1} \sigma_2^{A_2} \sigma_3^{A_3} .
\end{align}
Setting $T=\hat{1}+ \hat{2}+ \hat{3}$ and $\bar{B}=  T-B $, we have
\begin{align}
    \sigma^{( \bar{A})}=\sigma^{(T-A)}= \imag (-1)^{A_2} \sigma^{(A)}
\end{align}
and
\begin{align}
    \tr( {\sigma^{(A)}}^\dagger \sigma^{(B)}  ) = 2 (\delta_{AB} + \imag (-1)^{A_2}\delta_{A \bar{B}} ).
\end{align}
Then, $\chi$ is written as
\begin{align}
    \chi(2r+A)= \frac{1}{2}  \tr( {\sigma^{(A)}}^\dagger u(r) +\epsilon(A) {\sigma^{(A)}}^\dagger d(r) )
\end{align}
in terms of $u$ and $d$. Using the completeness relations
\begin{align}
\sum_{A} \sigma^{(A)}_{\alpha f} (\sigma^{(A)})^\dagger_{g\beta } = 4 \delta_{\alpha \beta} \delta_{fg} ,~\sum_{A} \sigma^{(A)}_{\alpha f} \epsilon(A) (\sigma^{(A)})^\dagger_{g\beta } =0,
\end{align}
the mass term becomes
\begin{align}
    \sum_{x} \chi(x)^\dagger \epsilon(x) \chi(x)= \sum_r \tr( d^\dagger(r) u(r) + u^\dagger(r) d(r) ).
\end{align}
On the other hand, the kinetic term is 
\begin{align}
   & \sum_{x} \eta_i(x) \chi^\dagger(x)( \chi(x+\hat{i})-\chi(x-\hat{i}) ) \nonumber\\
    = &   \sum_{r,A_i=0} \eta_i(A) \chi^\dagger(2r+A)( \chi(2r+A+\hat{i})-\chi(2(r-\hat{i}) +A+\hat{i}) ) \nonumber \\
     &+\sum_{r,A_i=1} \eta_i(A) \chi^\dagger(2r+A)( \chi(2(r+\hat{i})+A-\hat{i})-\chi(2r +A-\hat{i}) ).
\end{align}
The first is written by
\begin{widetext}
\begin{align}
    &\sum_{A_i=0} \eta_i(A) \chi^\dagger(2r+A) \chi(2r+A+\hat{i})\nonumber \\
    =& \sum_{A_i=0} (u(r) + \epsilon(A) d(r))^\dagger_{f \alpha} \sigma^{(A)}_{\alpha f} \eta(A) (\sigma^{(A+\hat{i})})^\dagger_{g \beta} (u(r)+\epsilon(A+\hat{i })d(r) )_{\beta g}  \nonumber\\
    =&\sum_{A_i=0} (u(r) + \epsilon(A) d(r))^\dagger_{f \alpha} \sigma^{(A)}_{\alpha f} \eta(A) (\sigma^{(A+\hat{i})})^\dagger_{g \beta} (u(r)-\epsilon(A)d(r) )_{\beta g} \frac{1+ (-1)^{A_i}}{2} \nonumber \\
    =& \frac{1}{2} \qty( u^\dagger(r)( \sigma^i \otimes 1) u(r) - d^\dagger(r) (\sigma^i \otimes 1) d(r) -u^\dagger(r) (1\otimes {}^t\sigma^i) d(r) +d^\dagger(r) (1\otimes {}^t\sigma^i) u (r) ).
\end{align}
\end{widetext}
The first term of the tensor product is a spin matrix, and the second one is a taste matrix.

\begin{comment}
\begin{align}
    \sum_{A_\mu=0} \eta_\mu(A) \chi^\dagger(2r+A) \chi(2(r-\hat{\mu})+A+\hat{\mu}) \\
    = \frac{1}{2} \qty( u^\dagger(r)( \sigma^\mu \otimes 1) u(r-\hat{\mu}) - d^\dagger(r) (\sigma^\mu \otimes 1) d(r-\hat{\mu}) -u^\dagger(r) (1\otimes {}^t\sigma^\mu) d(r-\hat{\mu}) +d^\dagger(r) (1\otimes {}^t\sigma^\mu) u (r-\hat{\mu}) )
\end{align}

\begin{align}
    \sum_{A_\mu=1} \eta_\mu(A) \chi^\dagger(2r+A) \chi(2(r+\hat{\mu})+A-\hat{\mu}) \\
    = \frac{1}{2} \qty( u^\dagger(r)( \sigma^\mu \otimes 1) u(r+\hat{\mu}) - d^\dagger(r) (\sigma^\mu \otimes 1) d(r+\hat{\mu}) +u^\dagger(r) (1\otimes {}^t\sigma^\mu) d(r+\hat{\mu}) -d^\dagger(r) (1\otimes {}^t\sigma^\mu) u (r+\hat{\mu}) )
\end{align}

\begin{align}
    \sum_{A_\mu=1} \eta_\mu(A) \chi^\dagger(2r+A) \chi(2r+A-\hat{\mu}) \\
    = \frac{1}{2} \qty( u^\dagger(r)( \sigma^\mu \otimes 1) u(r) - d^\dagger(r) (\sigma^\mu \otimes 1) d(r) +u^\dagger(r) (1\otimes {}^t\sigma^\mu) d(r) -d^\dagger(r) (1\otimes {}^t\sigma^\mu) u (r) )
\end{align}
\end{comment}

We calculate the others, and we have
\begin{comment}
\begin{align}
     \sum_{x} \eta_\mu(x) \chi^\dagger(x)( \chi(x+\hat{\mu})-\chi(x-\hat{\mu}) )= &\frac{1}{2} \left( u^\dagger (\sigma^\mu \otimes 1) (\nabla_\mu - \nabla_\mu^\dagger)  u -d ^\dagger (\sigma^\mu \otimes 1) (\nabla_\mu - \nabla_\mu^\dagger) d\right.  \\
     & \left. + u^\dagger(r) (1\otimes {}^t\sigma^\mu) (\nabla_\mu + \nabla_\mu^\dagger)d -d^\dagger(r) (1\otimes {}^t\sigma^\mu) (\nabla_\mu + \nabla_\mu^\dagger)u   \right).
\end{align}
\end{comment}
\begin{widetext}
\begin{align}
    H=& \frac{1}{2 }\sum_r  \Bigg[ u(r)^\dagger \sigma_i \otimes 1 \frac{\nabla_i - \nabla_i^\dagger}{2 \imag  } u(r) -d(r)^\dagger \sigma_i \otimes 1 \frac{\nabla_i - \nabla_i^\dagger}{2 \imag } d(r)   \nonumber\\
    &+ u(r)^\dagger 1\otimes {}^t\sigma_i \frac{\nabla_i + \nabla_i^\dagger  }{2 \imag } d(r)  -d(r)^\dagger 1\otimes {}^t\sigma_i \frac{\nabla_i + \nabla_i^\dagger  }{2\imag } u(r)  + 2 m ( u^\dagger(r) d (r) +d^\dagger(r) u(r) ) \Bigg],
\end{align}
\end{widetext}
where $\nabla_iu(x)=u(x+\hat{i}) -u(x) $ and $\nabla_i^\dagger u(r)= u(r-\hat{i})-u(r) $.

Let us define the $(d+1)$-dimensional spinor by 
\begin{align}
    \psi_f=  \mqty( u_f \\ d_f ).
\end{align}
This allows us to translate the Hamiltonian into the taste basis,
\begin{align}
H&= \frac{1}{2}\sum_r \psi^\dagger \Bigg[  ( \alpha^i \otimes 1 )\frac{\nabla_i - \nabla_i^\dagger }{2\imag } \\
&- ( \beta \gamma_5 \otimes {}^t \sigma^i )\frac{\nabla_i + \nabla_i^\dagger }{2\imag } +2 m ( \beta \otimes 1 )
\Bigg] \psi,
\end{align}
where 
\begin{align}
    \alpha^i&= \mqty(\sigma^i & 0 \\ 0 & -\sigma^i) ,~\beta=\mqty(0 & 1\\ 1 & 0 ) ,~ \nonumber\\
    \gamma_{5}&=-\imag \alpha^1 \alpha^2 \alpha^3= \mqty(1 & 0 \\ 0 & -1). 
\end{align}
The second term is regarded as a Wilson term, but mixes the flavors. In the continuum limit, this term gives doublers heavy mass, and the KS fermion system converges to $N_f= 2$-flavor Dirac fermion system. 

Note that, the shift operator $T_i$ acts on $u,~d$ and $\psi$ as 
\begin{align}
    T_i u(r) =& (\sigma_i \otimes 1) u(r)\nonumber \\
    &+ \frac{1}{2}  (\sigma_i \otimes 1)\nabla_i u(r) + \frac{1}{2}(1 \otimes {}^t \sigma_i )\nabla_i d(r),\\
    T_i d(r)=& - (\sigma_i \otimes 1 ) d(r) \nonumber \\
    &- \frac{1}{2}  (\sigma_i \otimes 1)\nabla_i d(r) - \frac{1}{2}(1 \otimes {}^t \sigma_i)\nabla_i u(r), \\
     T_i\psi(r)=&(\alpha^i \otimes 1) \psi(r) \nonumber \\
     &+  \frac{1}{2}( \alpha^i \otimes 1- \beta \gamma_5 \otimes {}^t \sigma^i ) \nabla_i \psi(r). 
\end{align}
Thus, it is equivalent to $\alpha^i \otimes 1$ in the continuum limit.

\subsection{Parity, Time Reversal, and Charge Conjugation}

Parity \eqref{eq: parity}, time reversal \eqref{eq: time reversal}, and charge conjugation \eqref{eq: charge conjugation} can be rewritten as 
\begin{align}
    P^{-1}\psi(r)P&= (\beta \gamma_5 \otimes 1)\Gamma^{-1} \psi(-r) \to (\beta \otimes 1)\psi(-r) , \\
    T^{-1}\psi(r) T &= T_1T_3 \psi(r) \to (\alpha_1 \alpha_3 \otimes 1) \psi(r), \\
    C^{-1}\psi(r) C&= (\beta\otimes 1) T_2 \psi^\ast(r) \to (\beta \alpha_2\otimes 1) \psi^\ast(r).
\end{align}
on the taste basis. We prove these equations in this section.

Parity changes $u$ to
\begin{align}
    P^{-1} u(r) P&=  P^{-1} \frac{1}{2} \sum_A \chi(2r+A) \sigma^{(A)}  P
    \nonumber \\
    &=\frac{1}{2} \sum_A  \chi(-2r-A ) \epsilon(A) \sigma^{(A)}.
\end{align}
Replacing $A$ with $T-A$, we have
\begin{align}
    P^{-1} u(r) P&=\frac{1}{2} \sum_A  \chi(-2r+A-T ) \epsilon(T-A) \sigma^{(T-A)}\nonumber \\
    &=-\frac{1}{2} \sum_A \imag (-1)^{A_2}  \chi(-2r+A-T) \epsilon(A) \sigma^{(A)} \nonumber \\
    &=-\Gamma^{-1} d(-r) .
\end{align}
Similarly, $d$ turns into
\begin{align}
  P^{-1}d(r)P&=  \frac{1}{2} \sum_A  \chi(-2r-A ) \sigma^{(A)}\nonumber \\
  &=\frac{1}{2} \sum_A  \chi(-2r+A-T ) \sigma^{(T-A)}\nonumber \\
  &= \Gamma^{-1} u(-r).
\end{align}
Then, parity acts on $\psi$ as
\begin{align}
    P^{-1} \psi(r) P=& \mqty( -\Gamma^{-1} d(-r) \\ \Gamma^{-1} u(-r)  )\nonumber \\
    =& \mqty( 0 & 1 \\ 1 & 0 ) \mqty( \Gamma^{-1} u(-r) \\ -\Gamma^{-1} d(-r)   )\nonumber \\
    =& (\beta \gamma_5\otimes 1) \Gamma^{-1} \psi(-r).
\end{align}

Next, time reversal transforms $u$ as
\begin{align}
    Tu(r)T^{-1}= &\frac{1}{2} \sum_A  T\chi(2r+A )T^{-1} (\sigma^{(A)})^{\ast}\nonumber \\
    =&\frac{1}{2} \sum_A  T_1 T_3\chi(2r+A ) \sigma^{(A)}\nonumber \\
    = &T_1 T_3 u(r).
\end{align}
This leads to 
\begin{align}
    T^{-1}\psi(r)T=   T_1 T_3  \psi(r).
\end{align}

Finally, charge conjugation for $u$ is given by
\begin{align}
    C^{-1} u(r) C=  &\frac{1}{2} \sum_A  (-1)^{A_2} \epsilon(A) T_2\chi^\ast(2r+A ) \sigma^{(A)}\nonumber \\
    = &\frac{1}{2} \sum_A   \epsilon(A) T_2\chi^\ast(2r+A ) (\sigma^{(A)})^\ast \nonumber \\
    =&T_2 d^\ast(r),
\end{align}
and we have
\begin{align}
    C^{-1}\psi(r) C= (\beta\otimes 1)T_2 \psi^\ast(r).
\end{align}

\subsection{Axial Flavor Symmetry}
On the taste basis, the axial flavor transformation $S_i$ can be translated as \cite{Golterman:2024xos}
\begin{align}
S_i \psi(r)& = (\gamma^5 \otimes {}^t \sigma_i) \psi(r)+ \frac{1}{2}( \gamma^5 \otimes {}^t \sigma_i -\beta \alpha_i\otimes 1) \nabla_i \psi(r) \nonumber \\
&\to  (\gamma^5 \otimes {}^t \sigma_i) \psi(r).
\end{align}

\subsection{Transformation by $Q_{S_T}$ on the Taste Basis}

We derive the action of $Q_{S_T}$ on the taste basis. At first, $Q_{S_T}$ acts on $u$ as
\begin{align}
    &[Q_{S_T}, u(r)]\nonumber\\
    &= \sum_A (-1)^{A_2} \qty(a(2r+A-T) -\imag b(2r+A+T)   )\sigma^{(A)}   \nonumber \\
    &=\sum_A (-1)^{A_2} \chi^\ast(2r+A+T) \sigma^{(A)}+ (\text{difference term}).
\end{align}
The first term becomes 
\begin{align}
    &\sum_A (-1)^{A_2} \chi^\ast(2r+A+T) \sigma^{(A)}\nonumber \\
    &= \sum_A \chi^\ast(2r+A+T) (\sigma^{(A) })^\ast \nonumber \\
    &=\sum_A \epsilon(A)  T_1 S_2 T_3 \chi(2r+A)^\ast (\sigma^{(A) })^\ast \nonumber \\
    &= T_1 S_2 T_3 d^\ast(r).
\end{align}
We perform the same calculation for $d$, and we have
\begin{align}
    [Q_{S_T} , \psi(r)]&= (\beta\otimes 1 )T_1 S_2 T_3 \psi^\ast(r) + (\text{difference term})\nonumber \\
    &\to -\imag \beta \alpha_2\otimes {}^t\sigma_2 \psi^\ast(r).
\end{align}
Note that the difference term can be ignored in the continuum limit.

\section{Vector and Axial Current on the Lattice}
\setcounter{equation}{0}
One may ask whether the conservation laws of the vector and axial currents are preserved on the lattice. In what follows, we derive the conservation laws associated with the vector $U(1)$ and axial $U(1)$ symmetries in lattice space, under the assumption that the field $\chi$ obeys the Schr\"odinger equation,
\begin{align}
    \pdv{}{t} \chi (x,t) =-\imag h\chi(x,t).
\end{align}

The vector $U(1)$ charge and current densities are defined as
\begin{align}
    j_V^0(x,t) &= \chi^\dagger(x,t)\, \chi(x,t), \\
    j_V^i(x,t) &= \frac{1}{2} \qty( \chi^\dagger(x,t)\, S_i \chi(x,t) 
    + (S_i \chi(x,t))^\dagger \chi(x,t) ).
\end{align}
These satisfy the lattice continuity equation,
\begin{align}
    \pdv{}{t} j_V^0(x,t) - \nabla^{\dagger}_i j_V^i(x,t) = 0,
\end{align}
where the difference operator $\nabla_i^\dagger$ is given by
\begin{align}
    \nabla_i^\dagger j_V^i(x,t) 
    = j_V^i(x-\hat{i},t) - j_V^i(x,t) .
\end{align}

We define the axial $U(1)$ charge and current densities as \cite{Golterman:2024xos}
\begin{align}
    j_A^0(x,t) &= \frac{1}{2} \qty( \chi^\dagger(x,t)\Gamma \chi(x,t) +h.c. ) , \\
    j_A^i(x,t) &= \frac{1}{4} \qty( \chi^\dagger(x,t)\, S_i \Gamma \chi(x,t) 
    + (S_i \chi(x,t))^\dagger \Gamma \chi(x,t) +h.c. ).
\end{align}
They satisfy the lattice continuity equation with an additional source term,
\begin{align}
    \pdv{}{t} j_A^0(x,t) - \nabla^{\dagger}_i j_A^i(x,t) 
    = \frac{1}{2}(\imag \chi^\dagger(x,t)\, [h,\Gamma]\, \chi(x,t) + h.c.)\label{eq: conservation law j_A}.
\end{align}
In particular, if $[h,\Gamma]=0$, the axial current is strictly conserved.

\section{Anomaly Equation in Continuum Spacetime}
\setcounter{equation}{0}
%\subsection{$(3+1)$-dimensional Dirac Fermion System}
\label{app: 3D continuum space}
In this section, we review Ref.~\cite{Dunne:1989gp}, and derive the chiral anomaly equation on the $(3+1)$-dimensional Dirac fermion system. The Hamiltonian is given by 
\begin{align}
    H=  \int_{\mathbb{R}^3} d^3x~\psi^\dagger h \psi(x), ~(h=  \alpha^i \qty(- \imag \partial_i + A_i)),
\end{align}
where $\alpha^i=\sigma_3\otimes \sigma^i$. We apply the homogeneous magnetic field in the $x^3$-direction:
\begin{align}
A_1=A_3=0,~A_2=Bx^1
\end{align}
for the positive constant $B>0$. The chiral operator is defined by
\begin{align}
    \gamma_5= -\imag \alpha^1 \alpha^2 \alpha^3= \sigma^3 \otimes 1.
\end{align}

We solve the eigenvalue problem of $h$. Since there are translational symmetries in $x^2$ and $x^3$ directions, we can take
\begin{align}
    \psi(x)= \frac{e^{\imag (k_2x^2+ k_3x^3)} }{2\pi} \xi_s \otimes \eta( x^1 ) .
\end{align}
Then, the Hamiltonian becomes
\begin{align}
   % h \psi (x)=  \frac{e^{\imag (k_2x^2+ k_3x^3)} }{2\pi} \xi_s \otimes s \mqty( k_3 & -\imag(\partial_1 +k_2+Bx^3   ) \\ -\imag(\partial_1 -k_2-Bx^3 ) & -k_3  )   \eta( x^1 ) .
    h \psi (x)=  \frac{e^{\imag (k_2x^2+ k_3x^3)} }{2\pi} \xi_s \otimes s \mqty( k_3 & a(k_2)  \\
   a^\dagger (k_2) & -k_3  )   \eta( x^1 ) ,
\end{align}
where
\begin{align}
    a(k_2)&=  -\imag(\partial_1 +k_2+Bx^3   ),~\\
    a^\dagger (k_2)&=  -\imag(\partial_1 -k_2-Bx^3 ),
\end{align}
and the commutation relation is given by
\begin{align}
    [a(k_2), a^\dagger (k_2)]=2B.
\end{align}
We find a normalizable mode $f(k_2+Bx^1)$ annihilated by $a(k_2)$:
\begin{align}
    f(k_2+Bx^1)= \exp(- \frac{1}{2}(k_2+Bx^1)^2 ) .
\end{align}
The normalization factor for $(a^\dagger(k_2))^n f $ is determined by
\begin{align}
    C_n^2=&\int dx^1 \abs{ (a^\dagger)^n f}^2\nonumber \\
    =& (2B)^n n! \int dx^1 e^{-(k_2+Bx^1)^2}\nonumber \\
    =&(2B)^n n! \frac{\sqrt{\pi }}{B}.
\end{align}
Setting
\begin{align}
    &\eta_+ ( n,k_2, k_3, x ) \nonumber \\
    &=\frac{1}{\sqrt{ (k_3+E)^2+ 2Bn }
    }\mqty( (k_3+ E) \frac{(a^\dagger)^{n-1} }{C_{n-1}}f \\ \sqrt{2Bn}  \frac{(a^\dagger)^{n} }{C_{n}}f ), \\
    &\eta_- ( n,k_2, k_3, x )\nonumber \\
    &=\frac{1}{\sqrt{ (k_3+E)^2+ 2Bn }} \mqty(   \sqrt{2Bn} \frac{(a^\dagger)^{n-1} }{C_{n-1}}f \\ -(k_3+ E) \frac{(a^\dagger)^{n} }{C_{n}}f )
\end{align}
for $n =1,2,\cdots $, and 
\begin{align}
     \eta ( n=0,k_2, k_3, x ) =\mqty( 0\\ \frac{1 }{C_{0}}f )
\end{align}
for $n=0$, we find the positive and negative energy states
\begin{align}
    u( s,  n,k_2, k_3, x  )=  &  \frac{e^{\imag (k_2x^2+ k_3x^3)} }{2\pi} \xi_s \otimes \eta_s(n,k_2,k_3,x^1), \\
    v( s,  n,k_2, k_3, x  )=  &  \frac{e^{\imag (k_2x^2+ k_3x^3)} }{2\pi} \xi_s \otimes \eta_{-s}(n,k_2,k_3,x^1)
\end{align}
for $n =1,2,\cdots $, and
\begin{align}
    &u( n=0,k_2, k_3, x  )\nonumber \\
    &=    \frac{e^{\imag (k_2x^2+ k_3x^3)} }{2\pi} \xi_{-\text{sign}(k_3) } \otimes \eta(n=0,k_2,k_3,x^1), \\
    &v( n=0,k_2, k_3, x  )\nonumber \\
    &=    \frac{e^{\imag (k_2x^2+ k_3x^3)} }{2\pi} \xi_{\text{sign}(k_3)} \otimes \eta(n=0,k_2,k_3,x^1)
\end{align}
for $n=0$. Let $\Omega =\Set{ (s,  n,k_2, k_3) }$ be a set of parameters characterizing the energy states. The normalizations are given by
\begin{comment}
\begin{align}
    \int dx^3~ u^\dagger (\Omega,x,t)  u(\Omega^\prime,x,t)=\int dx^3~ v^\dagger (\Omega,x,t)  v(\Omega^\prime,x,t)= \delta_{\Omega\Omega^\prime}, \\
    \int dx^3~ u^\dagger (\Omega,x,t)  v(\Omega^\prime,x,t)=0, \\
    \int d{\Omega}~ \qty[  u(\Omega,x,t) u^\dagger (\Omega,x^\prime,t) +v(\Omega,x,t) v^\dagger (\Omega,x^\prime,t)   ] = \delta_{xx^\prime},
\end{align}
\end{comment}
\begin{align}
\int d^3x\,
  u^\dagger(\Omega,x,t)\,u(\Omega',x,t)
&= \delta_{\Omega\Omega'}, 
\\
\int d^3x\,
  v^\dagger(\Omega,x,t)\,v(\Omega',x,t)
&= \delta_{\Omega\Omega'}, 
\\
\int d^3x\,
  u^\dagger(\Omega,x,t)\,v(\Omega',x,t)
&= 0,
\\
\int d\Omega\,
\Bigl[
  u(\Omega,x,t)\,u^\dagger(\Omega,x',t)\nonumber
 \\+ v(\Omega,x,t)\,v^\dagger(\Omega,x',t)
\Bigr]
&= \delta^{(3)}(x-x').
\end{align}

where
\begin{align}
    \int d\Omega = \int dk_2 dk_3 \sum_{n=1}^\infty \sum_{s=\pm} + \int dk_2 dk_3 \delta_{n,0}.
\end{align}
Then, the Dirac spinor $\psi$ can be expanded as 
\begin{align}
    \psi(x)= \int_{-\infty}^\infty dk [ b(k) u(k,x) + d^\dagger(k) v(k,x)  ],
\end{align}
where $b(k)$ and $d(k)$ are annihilation operators with momentum $k$. The commutator relations are given by
\begin{align}
    \qty{b(k), b^\dagger (k^\prime)}= \qty{d(k), d^\dagger (k^\prime)} \delta_{kk^\prime}
\end{align}
and all others are zero. We define the vacuum state such that it is annihilated by all $b(k)$ and $d(k)$:
\begin{align}
    b(k)\ket{0}= d(k)\ket{0}.
\end{align}

The massless Hamiltonian commutes with $\gamma_5$, which is called chiral symmetry. The chiral current is defined as
\begin{align}
    j_{A,c}^0(x)= &\lim_{y\to x} \psi^\dagger (x)\gamma^5 e^{-\imag A_i (y-x)^i } \psi(y), \\
    j_{A,c}^i(x)=& \lim_{y\to x} \psi^\dagger (x,t) \alpha^i \gamma^5 e^{-\imag A_i (y-x)^i } \psi(y,t)
\end{align}
with point splitting. Then, we have the vacuum expectation value \footnote{We regularize by inserting $e^{-\varepsilon|k_3|}$ and then take the limit $\varepsilon\to0^+$.}:
\begin{widetext}
\begin{align}
        \ev{j_{A,c}^0(x)}=&\lim_{y\to x}  \int d\Omega v^\dagger (\Omega, x) \gamma^5 v (\Omega, y) \nonumber\\
    =&\lim_{y\to x}  \int_{-\infty }^\infty dk^2 dk^3 \frac{e^{\imag ( k_2(y^2-x^2) + k_3(y^3-x^3) )}}{(2\pi)^2} \Bigg[  
    \text{sign}( k_3) \eta(0,k_2,k_3, x^1)^2 \nonumber  \\
    &\left.+\sum_{n=1}^\infty  ( \eta_{-}( n,k_2,k_3, x^1 )^2 - \eta_{+}( n,k_2,k_3, x^1 )^2  )  \right]   \nonumber \\
    =&\lim_{y\to x}  \int_{-\infty }^\infty dk^2 dk^3 \frac{e^{\imag    k_3(y^3-x^3) }}{(2\pi)^2}  \text{sign}( k_3) \frac{1}{C_0^2} e^{-(k_2+ Bx^1)^2} \nonumber \\
    =&\lim_{y\to x}   \frac{1}{(2\pi)^2}  B 2 \imag \frac{1}{y^3-x^3} = \lim_{y\to x} \imag  \frac{B}{ 2\pi^2}     \frac{1}{y^3-x^3}.
\end{align}
\end{widetext}
Using the rotational transformation, the general form is obtained by
\begin{align}
     \ev{j_{A,c}^0(x)}=\lim_{y\to x} \imag  \frac{B_i }{ 2\pi^2}  \frac{(y-x)^i}{ \norm{ y-x }^2 } .
\end{align}

Finally, we derive the anomaly equation by adding an adiabatic electric field. We carry out the same calculation as the previous section, and we have
\begin{align}
    \ev { \partial_\mu j_{A,c}^ \mu(x,t)}= \lim_{y\to x} \imag \dot{A}_i (y-x)^i \imag  \frac{B_j }{ 2\pi^2}  \frac{(y-x)^j}{ \norm{ y-x }^2 }.
\end{align}
Unlike the case of $(1+1)$ dimensional, there are many paths from $y$ to $x$, and the limit depends on the choice of them. According to \cite{Dunne:1989gp}, we should take the limit along the magnetic axis. That is, we set $y=\epsilon B+x $ and take the limit of $\epsilon \to 0$ as $\lim_{y\to x}$. Then, we find the anomalous divergence
\begin{align}
    \ev { \partial_\mu j_{A,c}^ \mu(x,t)}= - \frac{E_i B_i }{2\pi^2}
\end{align}
and 
\begin{align}
    \dv{}{t} \ev{Q_{A,c}(t) }= \dv{}{t} \int d^3x~ j_{A,c}^0(x,t)=- \int d^3x~  \frac{E_i B_i }{2\pi^2}.
\end{align}

\bibliographystyle{apsrev4-2}
\bibliography{ref}

\end{document}